\documentclass[pre,preprint,showpacs,aps,superscriptaddress]{revtex4-1}
\pdfoutput=1

\usepackage{graphicx}
\usepackage{amsmath}

\usepackage{amsmath}%
\usepackage{amsfonts}%
\usepackage{amssymb}%
\usepackage{enumitem}

\usepackage{textcomp}

\usepackage{color}
\definecolor{lasallegreen}{rgb}{0.03, 0.47, 0.19}
\definecolor{darkpastelgreen}{rgb}{0.01, 0.75, 0.24}
\definecolor{darkspringgreen}{rgb}{0.09, 0.45, 0.27}
\definecolor{emerald}{rgb}{0.31, 0.78, 0.47}
\definecolor{forestgreen(web)}{rgb}{0.13, 0.55, 0.13}
\definecolor{green(ryb)}{rgb}{0.4, 0.69, 0.2}
\definecolor{green(pigment)}{rgb}{0.0, 0.65, 0.31}
\definecolor{green(munsell)}{rgb}{0.0, 0.66, 0.47}

\newcommand{\bfF}{{\mathbf{F}}}

\newcommand{\bfJ}{{\mathbf{J}}}

\newcommand{\bfp}{{\mathbf{p}}}
\newcommand{\bfr}{{\mathbf{r}}}
\newcommand{\bfv}{{\mathbf{v}}}
\newcommand{\bfw}{{\mathbf{w}}}
\newcommand{\bfx}{{\mathbf{x}}}

\newcommand{\rmd}{{\mathrm d}}

\newcommand{\rms}{{\mathrm s}}

\newcommand{\cD}{{\mathcal D}}

\newcommand{\cJ}{{\mathcal J}}

\newcommand{\RR}{{\mathbb R}}

\newcommand{\TT}{{\mathbb T}}

\begin{document}

\def\Marseille{Aix-Marseille Universit\'e, UMR 7345 CNRS, 
Physique des interactions ioniques et mol\'eculaires, \\
campus Saint-J\'er\^ome, case 322, av.\ esc.\ Normandie-Niemen, FR-13397 Marseille cedex 20, France, EU}

\def\Recife{Departamento de F{\'\i}sica, Universidade Federal Rural de Pernambuco,
Rua Manoel de Medeiros, s/n - Dois Irm\~aos, 52171-900 - Recife, Brazil}

\def\Brasilia{Instituto de F{\'\i}sica and International Center for Condensed Matter Physics,
Universidade de Bras{\'\i}lia, CP 04455, 70919-970 - Bras{\'\i}lia, Brazil}

\def\BrasiliaICCMPh{International Center for Condensed Matter Physics,
Universidade de Bras{\'\i}lia, CP 04455, 70919-970 - Bras{\'\i}lia, Brazil}

\title{Contribution of individual degrees of freedom to Lyapunov vectors in many-body systems}

\author{L.~H.~Miranda Filho}
\email{lucmiranda@gmail.com}
\affiliation{\Marseille}
\affiliation{\Recife}
\affiliation{\Brasilia}
\author{M.~A.~Amato}
\email{maamato@unb.br}
\affiliation{\BrasiliaICCMPh}
\author{Y.~Elskens}
\email{yves.elskens@univ-amu.fr}
\affiliation{\Marseille}
\author{T.~M.~Rocha Filho}
\email{marciano@fis.unb.br}
\affiliation{\Brasilia}


\begin{abstract}

We use the weight $\delta I$, deduced from the estimation of Lyapunov vectors, in order to characterise regions in the kinetic $(x,v)$ space with  particles that most contribute to chaoticity. 
For the paradigmatic model, the cosine Hamiltonian mean field model, we show that this diagnostic highlights the vicinity of the separatrix, even when the latter hardly exists.


\par
\textit{Keywords} : chaos, Lyapunov exponents, Hamiltonian mean field, kinetic space, molecular dynamics.  
\end{abstract} 

\maketitle


\section{Introduction}
\label{sec.Intro}

Lyapunov exponents are part of the fundamental characterisation of dynamical systems \cite{ott, pikovsky}. 
They measure the rate of divergence between nearby trajectories due to small perturbation in the initial conditions, 
viz.\  they determine the chaoticity of dynamical systems. 
However, for many-body systems, they relate to the dynamics in the high-dimensional phase space of the dynamics, 
typically $\TT^{3N} \times \RR^{3N}$ for $N$ particles in a 3-dimensional cube with periodic boundary conditions. 

In addition, for identical particles, it is usual to describe the system's evolution 
not by a single point $(\bfr_1, \bfr_2, \ldots, \bfr_N, \bfp_1, \bfp_2, \ldots, \bfp_N)$,
but by an empirical \emph{measure} $N^{-1} \sum_{j=1}^N \delta (\bfr - \bfr_j) \delta(\bfv - \bfv_j)$ 
on the 6-dimensional \emph{kinetic space} $\TT^3 \times \RR^3$, viz.\ Boltzmann's $\mu$-space \cite{ehrenfest}.
For any finite $N$, the empirical measure and the single-point in Gibbs' $6N$-dimensional phase space 
provide equivalent information on the system microscopic state,
and their evolutions under the equations of motion are equivalent. 

This is the cornerstone of the mean-field derivation of the Vlasov equation in the $N \to \infty$ limit for smooth inter-particle interactions
\cite{spohn, EEDvlasovia13}. 
Considering that the kinetic description is physically illuminating about the behaviour of particles, we are tempted to search for a particle-related description of the chaoticity of the dynamics.
In this paper, we introduce a quantity characterising to what extent each particle, which moves in the kinetic space, 
contributes to the overall chaoticity in full phase space. 

To be specific, we consider a well-known mean-field system, the cosine Hamiltonian mean field model, 
in which particles move simply on a circle. The one-particle configuration space is the unit circle $\TT$, 
the $N$-particle configuration space is $\TT^N$, the one-particle kinetic space is $\TT \times \RR$, 
and the phase space is $(\TT \times \RR)^N$. 

In Sec.~\ref{sec.Lyap}, we provide a brief reminder on Lyapunov exponents and vectors for continuous-time systems. 
Sec.~\ref{sec.cosHMF} introduces the cosine Hamiltonian mean field model. 
Sec.~\ref{sec.num} presents the numerical method and our proposal to investigate sensitive regions in the kinetic space.
Sec.~\ref{sec.res} is devoted to our results, and Sec.~\ref{sec.conc} to our conclusion.

\section{Lyapunov Exponents}
\label{sec.Lyap}

Lyapunov exponents are an essential tool for discussing dynamical systems. They provide the distinction between regular and chaotic behaviour by measuring how small perturbations in the system evolve in phase space  \cite{abraham, alligood, bousko, dorfman, eckmann-ruelle, mane}.   
Let
\begin{equation}
  \bfx \equiv (x_1, x_2,....x_n)
\label{phase coord}
\end{equation}
be the coordinates in the $n$-dimensional phase space 
(for monatomic gas or liquid models one has  $n = 6N$ and $\bfx = (\bfr_i, \bfp_i), i=1, \ldots, N$ ), 
in which the system obeys autonomous first-order differential equations 
\begin{equation}
  \frac{\rmd \bfx(t)}{\rmd t} 
  = \bfF(\bfx(t)),
  \label{flow}
\end{equation}
generating a flow $\Phi(\bfx_0 ; t) = \bfx(t)$ from initial data $\bfx_0$ in this space. 
The vector field $\bfF(\bfx(t))$ is the velocity field of the flow. 
To measure contraction or stretching by $\bfF$ (over short times) and $\Phi$ (over long times) 
in the neighbourhood of the trajectory $\bfx(t)$ in phase space, 
consider the difference vector of two trajectories in this space, namely the deviation vector
\begin{equation}
  \bfw \equiv (\delta x_1, \delta x_2, \dots, \delta x_n)   .
  \label{eq.dev-vec}
\end{equation}
In the infinitesimal regime, the evolution equations for this deviation vector are the linear, first variation equations
\begin{equation}
  \frac{\rmd \bfw(t)}{\rmd t} 
  = \bfJ(\bfx(t)) \bfw (t) ,
  \label{eq.linear}
\end{equation}
with $\bfJ := \partial \bfF / \partial \bfx$ being the $n \times n$ Jacobian matrix of the vector field. 
These equations also generate the evolution of the Jacobian matrix of the flow 
$\cJ(\bfx_0 ; t) = \left.\frac{\partial \Phi(\bfx_0 ; t)}{\partial \bfx_0}\right|_t$,
\begin{equation}
  \frac{\rmd \cJ(\bfx_0 ; t)}{\rmd t} 
  = \bfJ(\bfx(t)) \cJ(\bfx_0 ; t) .
\label{eq.jacflow}
\end{equation}

The evolution equations ($\ref{eq.linear}$) are integrated with an initial condition $\bfw_0 = \delta \bfx(0)$. 
When the elements of $\bfJ$ are continuous bounded functions of $t$, 
the solutions of (\ref{eq.linear}) grow no faster than $\exp(\gamma t)$, for some finite $\gamma$.
The Lyapunov exponents are defined by
\begin{equation}
  \lambda (\bfx_0, \bfw_0)
  = \displaystyle \lim_{t \to \infty} \frac{1}{t} \ln \frac{\| \bfw(t) \|}{\| \bfw_0 \|}
\label{eq.def-Lyap}
\end{equation}
when the limit exists. 
A priori, this limit depends on both the initial data $\bfx_0$ and the initial deviation $\bfw_0$
since $\bfJ(\bfx(t))$ depends on the trajectory and $\bfw_0$ is the initial data to solve (\ref{eq.linear}). 
Thus, in an $n$-dimensional system, one has $n$ Lyapunov exponents and each of them refers to the divergence degree of the characteristic directions of the system. 
All these exponents, with their multiplicity, form the Lyapunov spectrum, which is usually ordered as 
\begin{equation}
  \lambda_1 \geq \lambda_2 \geq \dots \geq \lambda_n.  
\label{eq.LS}
\end{equation}
The Largest Lyapunov Exponent (LLE) is denoted by $\lambda_1$. The existence of  divergent trajectories and therefore chaotic regime may be uncovered by positive values of the exponents.

When the trajectory of interest $\bfx(t)$ visits (densely) an extended domain $\cD$ in phase space,
the Lyapunov spectrum does not depend specifically on $\bfx_0$ and simply characterises 
this domain, which is foliated by manifolds associated with the characteristic directions singled out by the long-time behaviour of the $\bfw_i$'s. 
The asymptotic $\bfw_i$'s (up to their normalisation) are the associated Lyapunov vectors \cite{posch}. 

Hamiltonian dynamics are described in terms of pairs of conjugate variables and the form of Hamilton's equations ensures to the Lyapunov spectrum a special symmetry property, i.~e., the Lyapunov exponents come in pairs \cite{pikovsky}
\begin{equation}
  \lambda_i = - \lambda_{n-i+1}.
\label{eq.LS_H}
\end{equation}


In the following sections, we introduce the model, the cosine Hamiltonian mean field model, and the numerical approach to calculate the Lyapunov exponents. Along with these, we also discuss the numerical instruments used to investigate the sensitivity of specific degrees of freedom 
to the characteristic exponential behaviour associated with the largest Lyapunov exponent. 
This is built into the definition of the characteristic vectors, which may have arbitrary directions in phase space~:
our diagnostic attempts at identifying those regions of the kinetic space where particles contribute most to chaoticity. 
Particles in those regions are also those which will be most important 
when one attempts to reduce the many-body dynamics to a smaller number of degrees of freedom \cite{antoniazzi}.

\section{The cosine Hamiltonian mean field model}
\label{sec.cosHMF}

We now briefly introduce the $N$-body model used in this paper. 
The model has a Hamiltonian form
\begin{equation}
  H 
  = \sum_{\ell=1}^{N} \frac{{p_\ell}^2}{2m} 
     + \frac{1}{N} \sum_{\ell=1}^{N-1} \sum_{j=\ell+1}^N V({r_\ell} - {r_j})   ,
\label{ham_hmf}
\end{equation}
where $m$, $p_\ell$ and $r_\ell$ are the mass, momentum and position of particle $\ell$, respectively. 
The $1/N$ factor is Kac' factor which makes the total energy an extensive quantity \cite{kac-meanfield} 
and ensures that the dynamics of individual particles is well defined in the large $N$ limit under mild conditions 
\cite{spohn, trocheris, EEDvlasovia13,kie14}.

The cosine Hamiltonian mean field (cosHMF) is a toy model consisting of 
$N$ classical particles moving on a unit circle and interacting via long range force~\cite{AntoniRuffo}. 
It is a widely studied system in the literature as it is solvable at equilibrium and allows for fast molecular dynamics simulations as they scale with the number $N$
of particles instead of $N^2$ for more generic systems. 
Moreover, analytical results for the scaling behaviour of the largest Lyapunov exponent with $N$ were obtained by Firpo \cite{firpo1998} and recently approached numerically by \cite{miranda2018}.

The particle masses $m$ and coupling constant of the interaction can be eliminated by a simple rescaling, 
so that $p$, $\theta$, $H$, time and Lyapunov exponents are dimensionless. 
The interaction energy is given by
\begin{equation}
  V(\theta_\ell - \theta_j) 
  = 1 - \cos{(\theta_\ell - \theta_j)},
\label{hmf}
\end{equation}
with the angles $\theta_\ell$ representing the coordinates $r_\ell$ of the particles. 
In this paper, we only consider the attractive case. 
The equations of motion generated by \eqref{ham_hmf},
\begin{align}
  \frac{\rmd \theta_\ell}{\rmd t} 
  & = p_\ell     ,
  \label{eq.thetadot}
  \\
  \frac{\rmd p_\ell}{\rmd t}
  & = - M \sin ( \theta_ \ell - \varphi)  = - M_x \sin \theta_\ell + M_y \cos \theta_\ell    ,  
  \label{eq.pdot} 
\end{align}
are such that each particle feels the effect of all other ones only through the value of the magnetisation (mean field) 
\begin{equation}
  \vec{M} 
  = (M_x, M_y) 
  = (M \cos \varphi, M \sin \varphi)
  = \frac{1}{N} \sum_{\ell=1}^{N} (\cos{\theta_\ell}, \sin{\theta_\ell}) ,
\label{mag}
\end{equation}
which is the first Fourier component of the particles spatial density. 

This implies that each particle behaves like a charged particle 
in the field of a wave with amplitude $M= | \vec M |$ and phase $\varphi = {\mathrm{Arg}}\ \vec M$. 
When the magnetisation is almost constant, particles are almost independent, 
though they are a priori coupled through their individual contributions to $\vec M$ via (\ref{mag}). Then, it makes sense to analyse their motion in the kinetic-theoretical $(\theta, p)$ space 
where each particle may appear either trapped in the potential well defined by $\vec M$, 
or circulating with respect to this well, viz.\ in terms of a reduced one-particle effective Hamiltonian
\begin{equation}
  h_1 (\theta', p' ; \vec M, \dot \varphi) 
  =
  \frac{1}{2} (p' - \dot \varphi)^2 + M (1 - \cos (\theta' - \varphi))
  \label{eq.h1}
\end{equation}
which generates \eqref{eq.thetadot}-\eqref{eq.pdot} for $(\theta_\ell, p_\ell) = (\theta', p')$ 
after a Galileo transformation to the wave frame when $\dot \varphi$ is constant. 

As the wave moves at its own velocity $\dot \varphi$, it is sensible to understand this trapping in the reference frame of its well \cite{EEbook}. 
The border between trapping and circulating behaviour occurs at $h_1 = e_\rms$ with the separating energy $e_\rms = 2 M$ in the wave frame,
and it is drawn in the one-particle $(\theta, p)$ space by the two branches of the separatrix, 
with equation 
\begin{equation}
  p - \dot \varphi 
  = 
  \pm 2  \sqrt{M}  \cos \frac{\theta - \varphi}{2}    .
  \label{eq.separ}
\end{equation}
It is well-known that the pendulum has regular dynamics, 
but trajectories on either side of the separatrix diverge exponentially in time (hence the name separatrix) \cite{EEbook}.

Note that the dynamics \eqref{eq.thetadot}-\eqref{eq.pdot} preserves total energy $H = \sum_\ell p_\ell^2 / 2 + (1 - M^2) N/2$ 
and total momentum $ P = \sum_\ell p_\ell = \sum_\ell \dot \theta_\ell$. 
It does not conserve the magnetisation, as 
\begin{align}
  \dot M_x 
  &= - N^{-1} \sum_\ell \dot \theta_\ell \sin \theta_\ell    ,
  \label{eq.Mxdot}
  \\
  \dot M_y 
  &= N^{-1} \sum_\ell \dot \theta_\ell \cos \theta_\ell    .
  \label{eq.Mydot}
\end{align}
Actually, for finite values of $N$, the magnetisation cannot be exactly conserved by the dynamics 
unless it vanishes exactly \cite{elskensESAIM}.

Note also that the total energy $H$ bears no simple relation with the reduced one-particle energies for individual particles,
for two reasons. First, $h_1$ is more conveniently expressed in the magnetisation comoving frame.
Second, even when $\dot \varphi = 0$, the potential term in $h_1$ sums essentially to twice the total potential energy in $H$, 
while the kinetic term in $h_1$ sums to the total kinetic energy in $H$. 
In other words, the average of $h_1$ over all particles is not the average energy per particle $e := H/N$. 

The equilibrium statistical mechanics of system \eqref{ham_hmf}-\eqref{hmf} can be solved analytically 
in both canonical and microcanonical ensembles \cite{AntoniRuffo}. 
Its caloric curve shows a second order phase transition at the critical energy per particle $e^\ast = E^\ast / N = 3/4$.
In terms of the order parameter $|\vec M|$, the critical energy separates two types of solutions, 
those with $|\vec M| \neq 0$ ($e^\ast<3/4$) and $|\vec M| = 0$ ($e^\ast > 3/4$). 
We refer to these regimes as subcritical and supercritical, respectively.

\section{Numerical method}
\label{sec.num}

By definition, the calculation of the Lyapunov exponents requires 
the knowledge of how the flow $\Phi(\bfx_0 ; t)$, generated by (\ref{flow}), 
and its local deviations $\{\delta x_i\}$, evolving by (\ref{eq.linear}), behave. Starting from a set of orthogonal deviation vectors, they are periodically re-orthonormalised at the same time that the expansion of volumes of different dimensions is computed. 

This normalisation rescaling prevents the divergence of the deviation vectors $\delta\bfx(t)$, and the orthogonalisation ensures the calculation of rate of divergence along linearly independent directions in space. The standard numerical approach is available in the references \cite{benettin1980,shimada1979,wolf1985}   

The perturbations evolve in the tangent space, obeying the linearised equations of motion \cite{benettin_1976,parker}. 
The appropriate equations are 
\begin{eqnarray}
  \begin{array}{ll}
  \bfx \equiv (x_1, x_2, \dots, x_n), 
  & \dot{\bfx}(t) = \bfF(\bfx(t))  ,
  \\
  \bfw_i \equiv (\delta x_{i,1}, \delta x_{i,2}, \dots, \delta x_{i,n}), 
  & \dot{\bfw}_i (t) \,= \bfJ(\bfx(t)) \bfw_i (t) \,, \hspace{1.00 cm} 1 \leq i \leq n 

\end{array}
\label{mot_eq}
\end{eqnarray}
where one may set $\| \bfw_i(0) \| = 1$ for convenience in the Lyapunov exponents calculation. 
The Jacobian (evaluated along the reference trajectory) and $\bfw_i(t)$ 
refer to the linearised versions of the actual equations of the system. 
Taking an initial condition $\bfx_0$ 
and an orthonormal basis $( \pmb{\delta}^{(0)}_{1}, \pmb{\delta}^{(0)}_{2}, \dots, \pmb{\delta}^{(0)}_{n})$ for the linear equations, 
we carry out the numerical integration of equations \eqref{mot_eq}. 
In this process, one has a single nonlinear equation and a set of $n$ linearised equations. 
Every new point $\bfx$ in phase space provides a different Jacobian matrix, 
so that each time step of integration involves a new linear operator $\bfJ$. 
Since $\| \bfw_i(t) \|$ diverges exponentially (with time), 
we apply at every period $T$ the Gram-Schmidt orthogonalisation procedure 
to the basis $(\bfw_1(t), ... \bfw_n(t))$. 
This process is iterated $K$ times to estimate the Lyapunov spectrum and vectors. 

The steps for obtaining the exponents are thus~:
\begin{enumerate}[label=\roman*)]
\item {\small for the nonlinear system, choose an initial condition $\bfx_0$~; 
for the $n$ linearised equations, define an orthonormal set of initial conditions 
$( \pmb{\delta}^{(0)}_{1}, \pmb{\delta}^{(0)}_{2}, \dots, \pmb{\delta}^{(0)}_{n})$~;}
\item {\small the whole set of $1+n$ differential equations is integrated simultaneously for a lapse $T$: $\bfx_0 \mapsto \bfx(T)$ and 
$( \pmb{\delta}^{(0)}_{1}, \pmb{\delta}^{(0)}_{2}, \dots, \pmb{\delta}^{(0)}_{n}) 
\mapsto ( \pmb{\delta}^{(1)}_{1}, \pmb{\delta}^{(1)}_{2}, \dots, \pmb{\delta}^{(1)}_{n})$~;}
\item {\small intervene on the solutions of linearised equations with the Gram-Schmidt procedure~;
the orthogonalisation 
$(\pmb{\delta}^{(1)}_{1}, \pmb{\delta}^{(1)}_{2}, \dots, \pmb{\delta}^{(1)}_{n})
\mapsto (\pmb{v}^{(1)}_{1}, \pmb{v}^{(1)}_{2}, \dots, \pmb{v}^{(1)}_{n})$ 
uses a triangular matrix with unit entries on its diagonal,  
while the normalisation generates rescalings $(\pmb{v}^{(1)}_{1}, \pmb{v}^{(1)}_{2}, \dots, \pmb{v}^{(1)}_{n})
\mapsto (\pmb{u}^{(1)}_{1}, \pmb{u}^{(1)}_{2}, \dots, \pmb{u}^{(1)}_{n})$~;}
\item  {\small take the updated basis $(\pmb{u}^{(1)}_{1}, \pmb{u}^{(1)}_{2}, \dots, \pmb{u}^{(1)}_{n})$ 
and $\bfx(T)$ as new initial conditions, 
for the linearised and nonlinear equations respectively, 
and repeat the process $K$ times~;}
\item {\small the Lyapunov exponents are estimated as time averages 
$\lambda_i = \displaystyle \frac{1}{KT} \sum_{k=1}^K \ln ||\pmb{v}^{(k)}_{i}||$. 
Stop iterating when $K$ is large enough to get convergence in the values of $\lambda_i$'s.}
\end{enumerate}  

In particular, the Lyapunov exponents calculation for cosHMF model leads to the deviation vector
$\bfw \equiv (\delta \theta_1, \delta \theta_2, \dots, \delta \theta_N, \dots, \delta p_1, \delta p_2, \dots, \delta p_N)$,
associated to the linearised equations 
\begin{align}
  \frac{\rmd \delta \theta_\ell}{\rmd t} 
  & = \delta p_\ell     ,
  \label{eq.dthetadot}
  \\
  \frac{\rmd \delta p_\ell}{\rmd t}
  & =    -M\cos (\theta_\ell - \varphi) \delta \theta_\ell + \frac{1}{N}\sum_{j=1}^{N}\cos(\theta_j - \theta_\ell) \delta\theta_j,
  \label{eq.dpdot} 
\end{align}
that describe the fluctuations evolution $\bfw$ 
around $\bfx \equiv (\theta_1, \theta_2, \dots, \theta_N, p_1, p_2, \dots, p_N)$.

With unit probability, an arbitrary initial perturbation $\bfw_0$ will generate a vector converging to the vector $\pmb{u}_1$ in phase space,
associated with the largest Lyapunov exponent. 
We attribute to each particle the weight 
\begin{equation}
  \delta I_\ell = \frac{(\delta\theta_\ell)^2 + (\delta p_\ell)^2}{|\bfw|^2}  
  \label{deltaI_def}
\end{equation}
to estimate its individual participation to the divergence rate between nearby trajectories in $2N$-dimensional phase space.  
This weight (normalised to unity) can be interpreted as the sensitivity level 
of the region where particle $k$ is in the kinetic space, 
which leads to the operative notion of Lyapunov modes \cite{Borzsak96,mila98,taniguchi2003B,bosetti10,ginelli11,bosetti14}.

\section{Results}
\label{sec.res}

\begin{figure}[!h]
\begin{center}
\begin{minipage}[b]{0.49\linewidth}
\includegraphics[width=\linewidth]{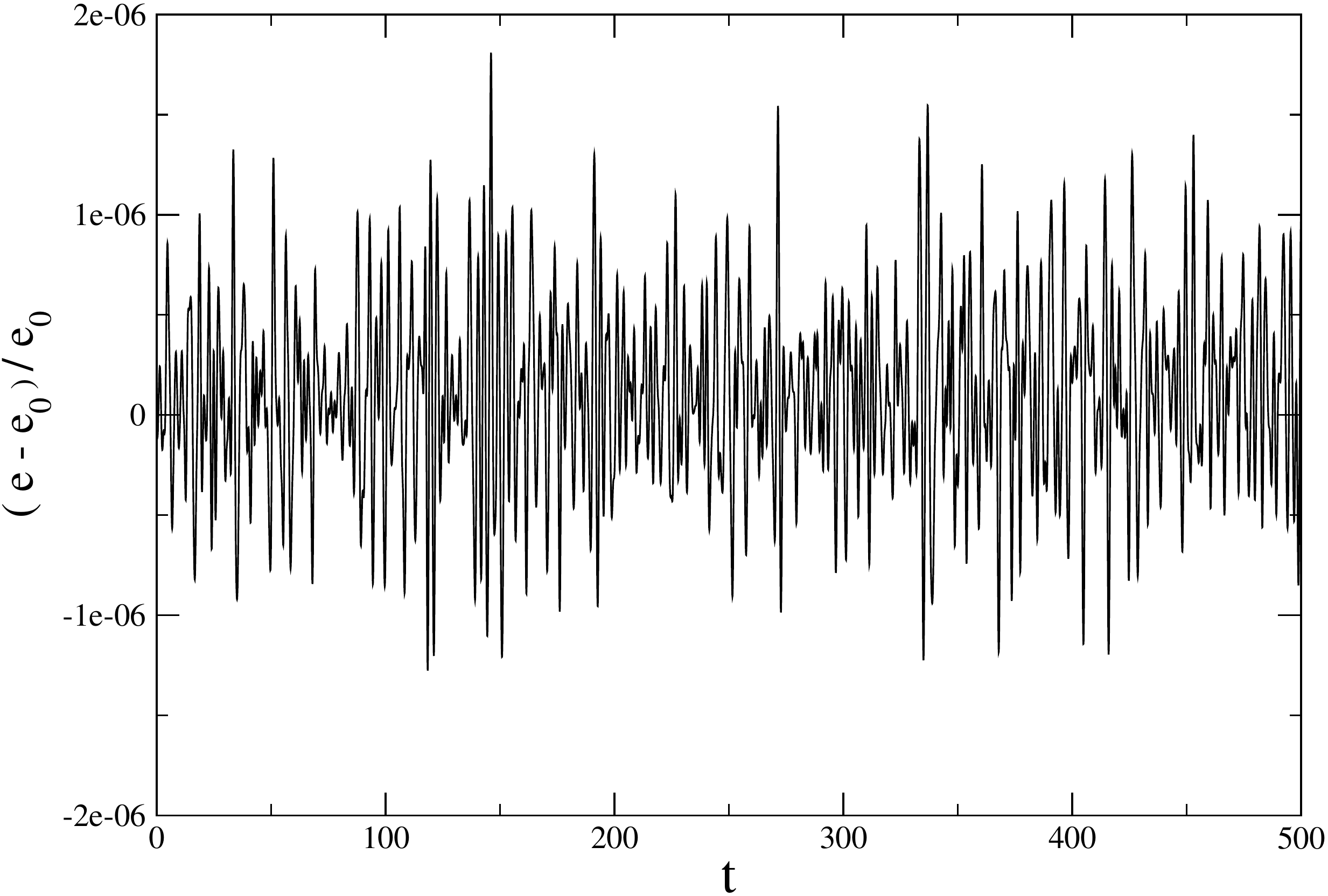}
\end{minipage} \hfill
\begin{minipage}[b]{0.49\linewidth}
\includegraphics[width=\linewidth]{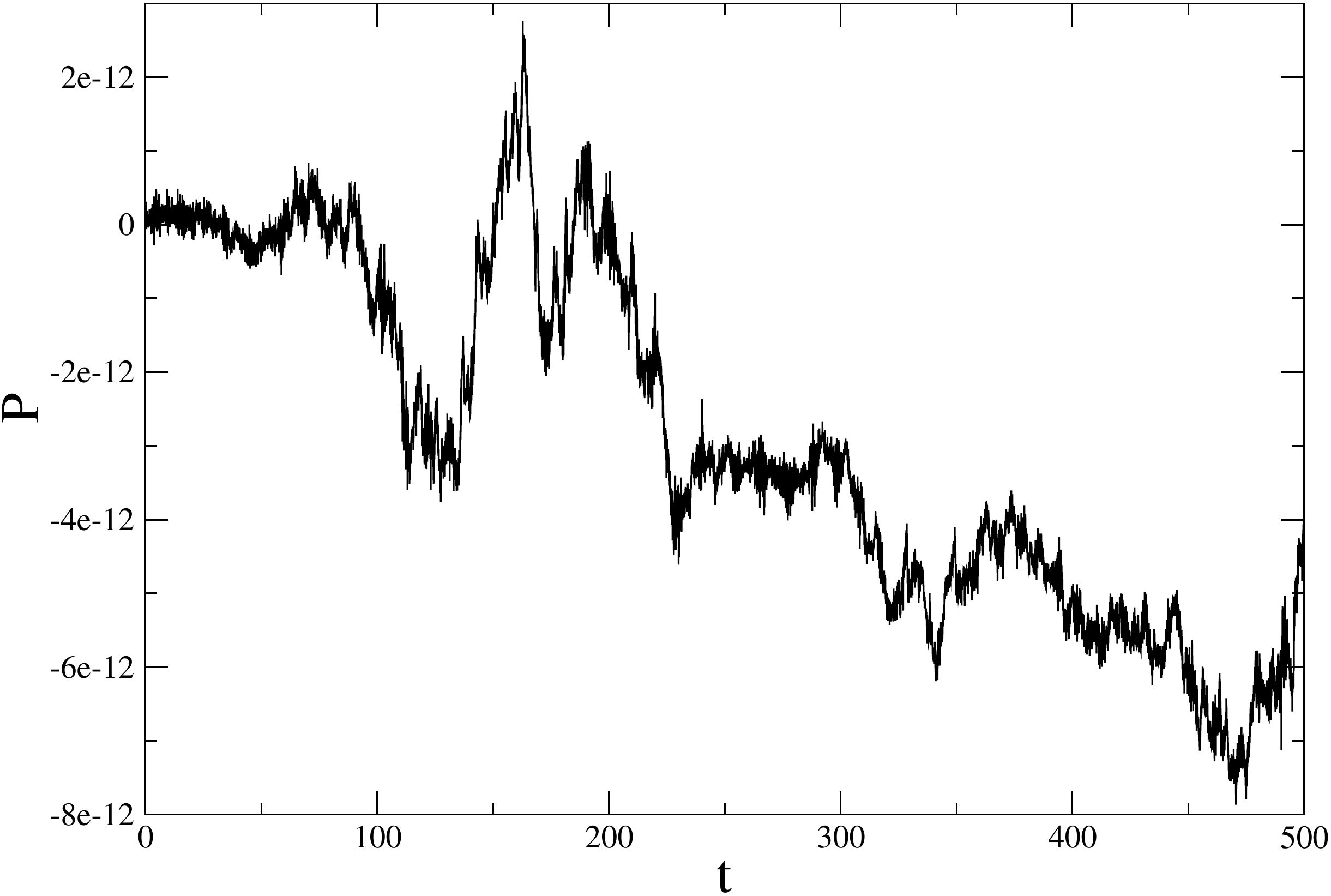}
\end{minipage} \hfill
\caption{$N = 10^4$, $e \approx 0.55$. Relative error of the energy and total momentum versus time.}
\label{erro_HMF}
\end{center}
\end{figure}

The evolution equations (\ref{eq.dthetadot})-(\ref{eq.dpdot}) are solved numerically 
using the centered leapfrog method with a time step $\Delta t = 0.\,05$. 
We need a symplectic integrator to ensure that 
the numerical solution preserves the Hamiltonian nature of the model. 
Moreover, the method is time-reversal invariant, second-order and robust \cite{hairer}. 
Figure~\ref{erro_HMF} shows the fluctuations in the energy \eqref{ham_hmf} and the total momentum $P$.

\begin{figure}[!h]
\begin{center}
\begin{minipage}[b]{0.45\linewidth}
\includegraphics[width=\linewidth]{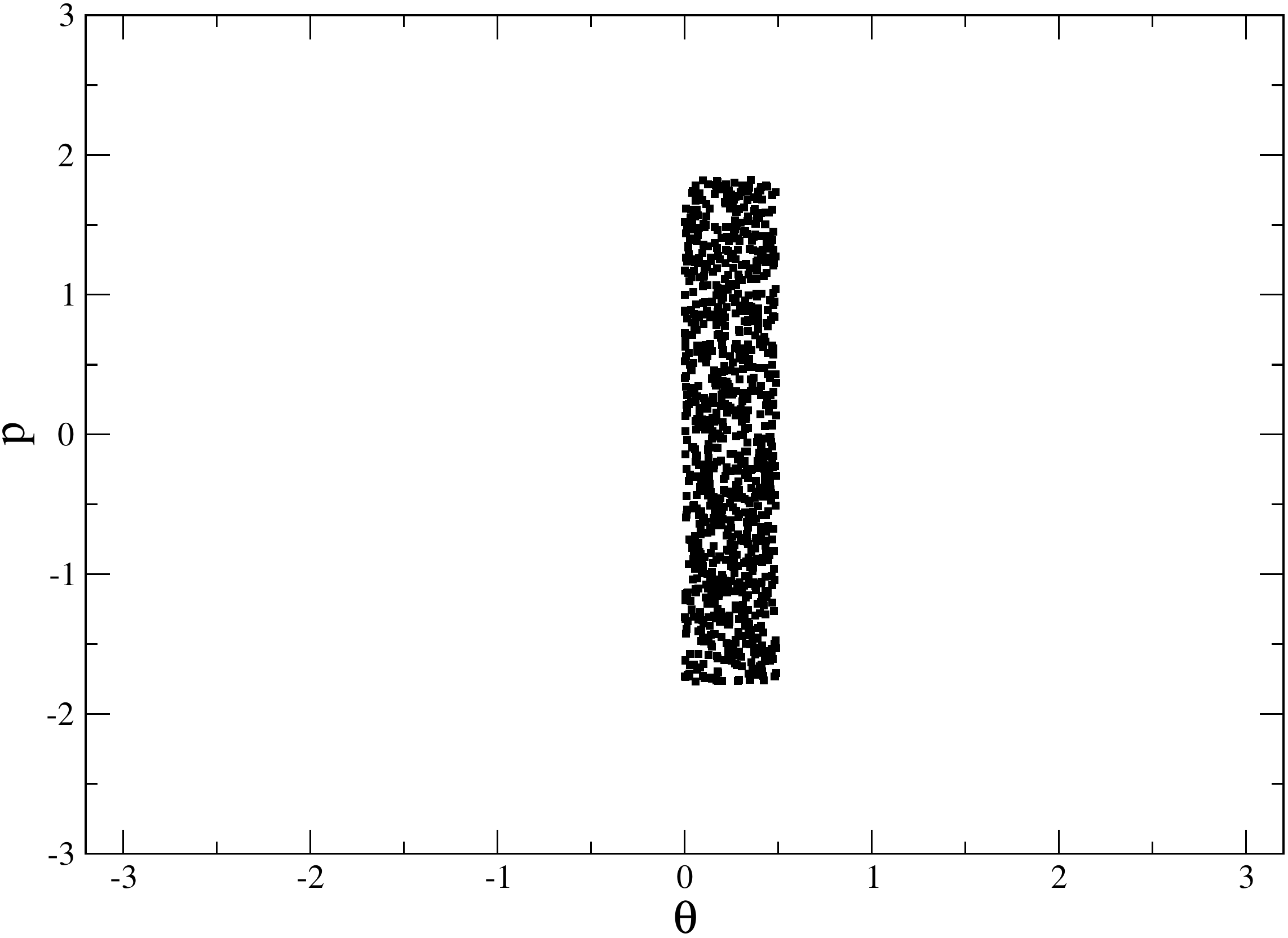}
\end{minipage} \hfill
\begin{minipage}[b]{0.48\linewidth}
\includegraphics[width=\linewidth]{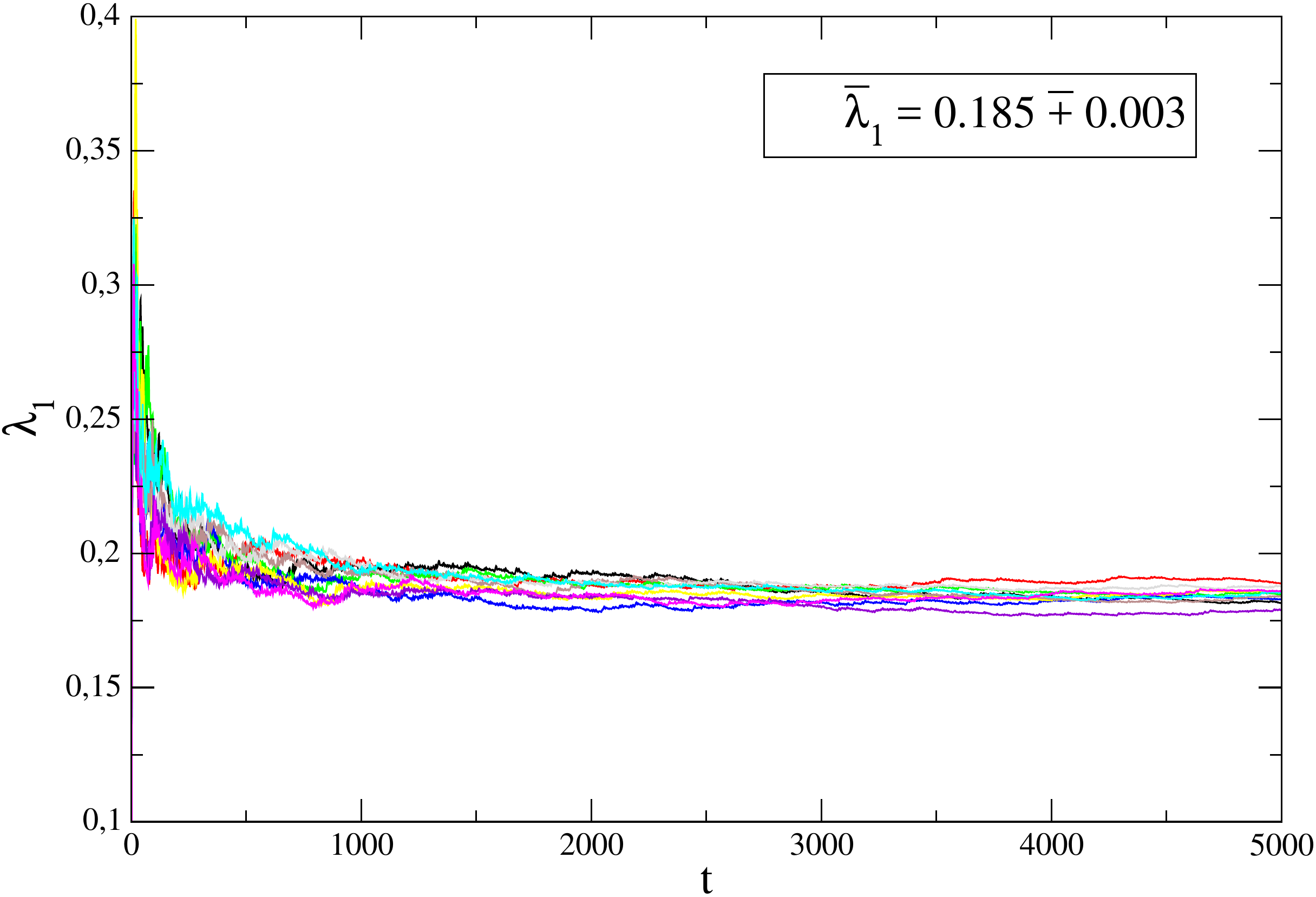}
\end{minipage} \hfill
\end{center}
\caption{$N = 10^3$, $e \approx 0.\,55$. (Left) Initial waterbag distribution with $\Delta \theta = 0.\,49$ and $\Delta p/2 = 1.\,80$ that corresponds to the energy $e \approx 0.\,55$. (Right) Largest Lyapunov exponent value computed for 10 realisations versus simulation time.}
\label{conv_wb}
\end{figure}

\begin{figure}[!h]
\begin{center}
\begin{minipage}[b]{0.45\linewidth}
\includegraphics[width=\linewidth]{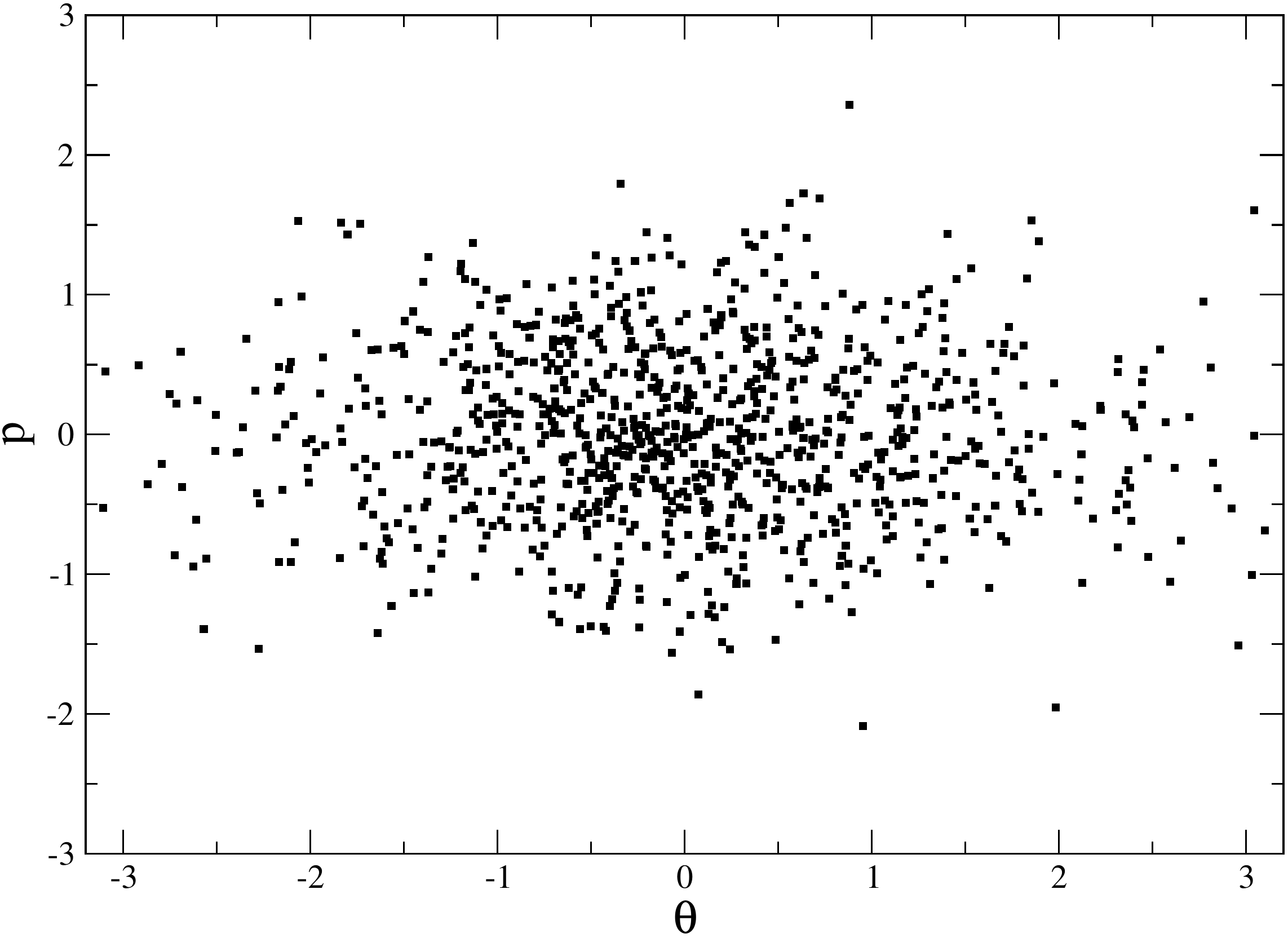}
\end{minipage} \hfill
\begin{minipage}[b]{0.48\linewidth}
\includegraphics[width=\linewidth]{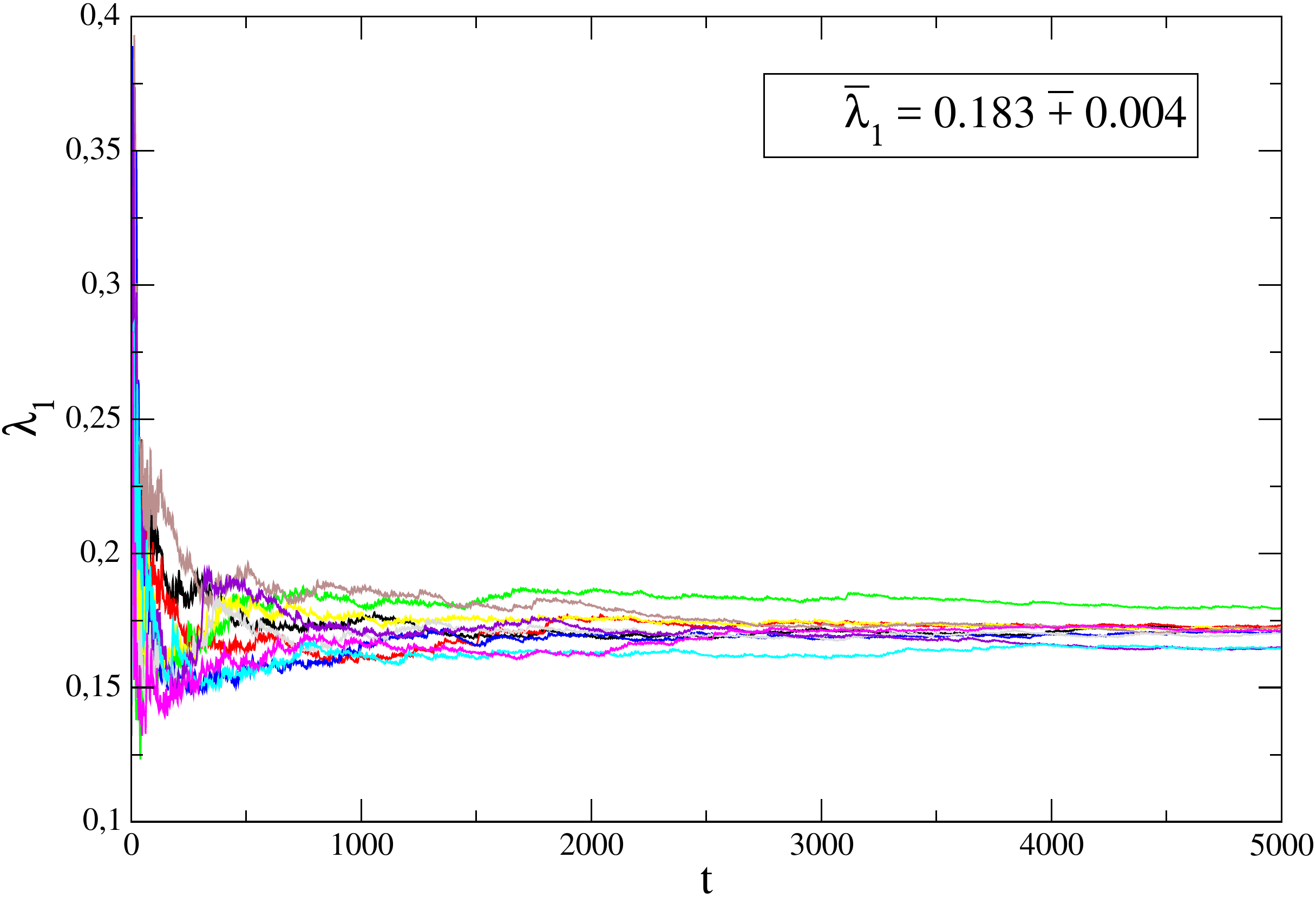}
\end{minipage} \hfill
\end{center}
\caption{$N = 10^3$, $e \approx 0.\,55$. (Left) Initial equilibrium distribution with Gaussian momentum distribution, kinetic energy per particle $\approx 0.\,21$, initial magnetisation $|\vec M_0| \approx 0.\,55$, corresponding to the cosHMF equilibrium with energy per particle $e \approx 0.\,56$. (Right) Largest Lyapunov exponent value computed for 10 realisations versus simulation time.}
\label{conv_eq}
\end{figure}

Numerical techniques described in previous section are applied to the cosHMF model in order to obtain the Lyapunov exponents. 
Figures~\ref{conv_wb} and \ref{conv_eq} show in details the  convergence of the numerical procedure
and the respective initial conditions for which the equations of motion~(\ref{eq.thetadot})-(\ref{eq.pdot}) are solved. 

The initial condition displayed in Figure~\ref{conv_wb} corresponds to an out of equilibrium configuration called waterbag.
This distribution is defined as $f(\theta,p) = 1/(\Delta \theta \Delta p)$ 
if $0 < \theta < \Delta \theta$ and $|p| < \Delta p/2$, and $f(\theta,p) = 0$ otherwise. 
This random uniform distribution over the rectangle of area $\Delta \theta \times \Delta p$
provides a state in which the magnetisation is $M_0 = [(1-\cos{\Delta \theta})^2+(\sin{\Delta \theta})^2]^{1/2} /\Delta \theta$ and the energy is $e = \Delta p^2/24 + (1 - M_0^2)/2$ \cite{rocha-amato-1}.   

Figure~\ref{conv_eq} is obtained from equilibrium initial data. 
In this case, the initial magnetisation and momentum 
are determined according to solutions of equilibrium statistical mechanics    
and generated randomly by Monte Carlo method \cite{davidp}.

As mentioned previously, the growth of the solutions to (\ref{eq.linear}) is bounded, and, for the same energy and number of particles, 
the Lyapunov exponents should not depend on a particular trajectory around which the perturbation is defined. To this end, one runs $10$ realisations for both types of initial conditions and show the convergence of $\lambda_1$. 
From here on, our results are generated from the equilibrium initial data as in Figure~\ref{conv_eq}.       

\begin{figure}[!h]
\begin{center}
\begin{minipage}[b]{0.49\linewidth}
\includegraphics[width=\linewidth]{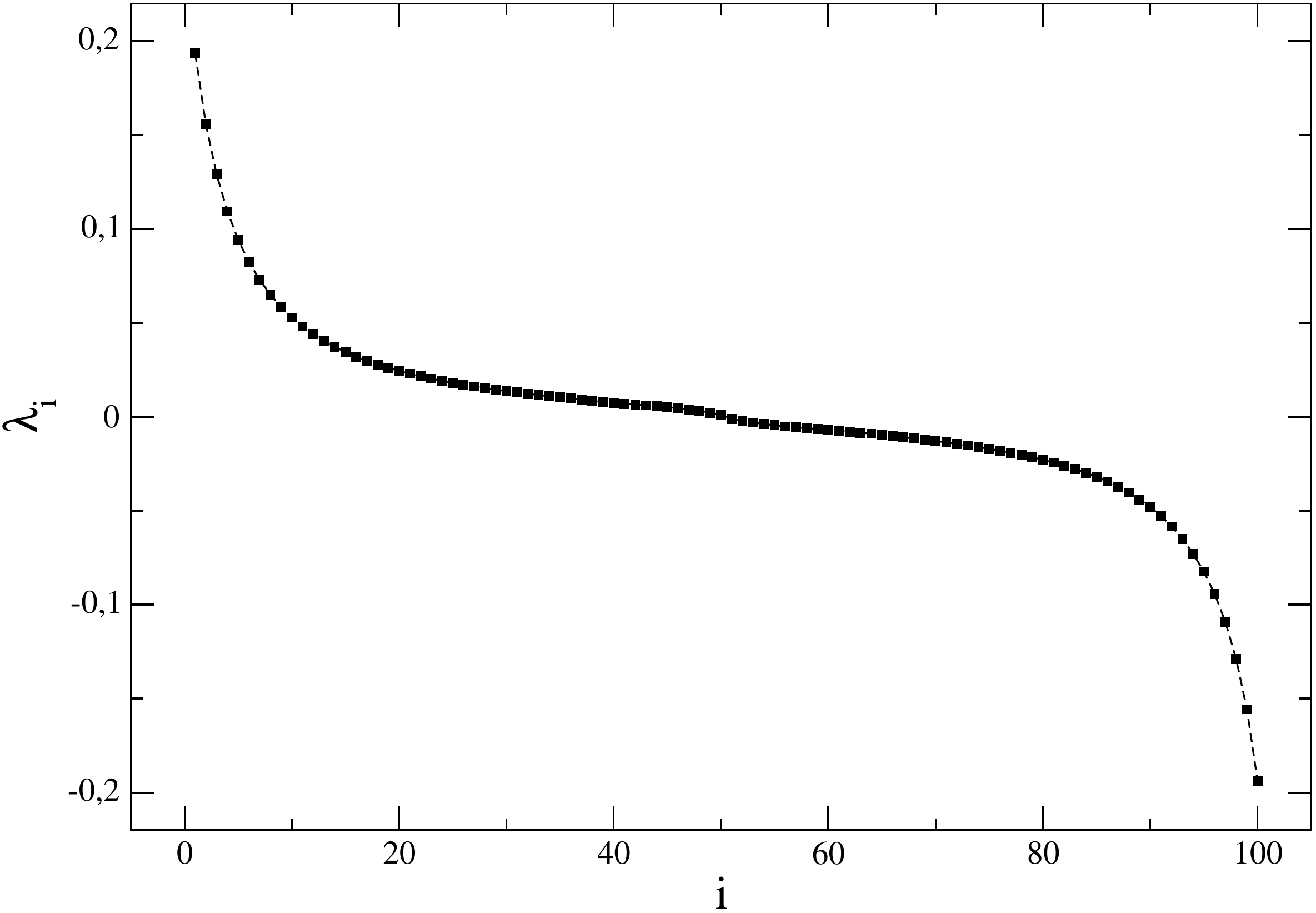}
\end{minipage}
\begin{minipage}[b]{0.49\linewidth}
\includegraphics[width=\linewidth]{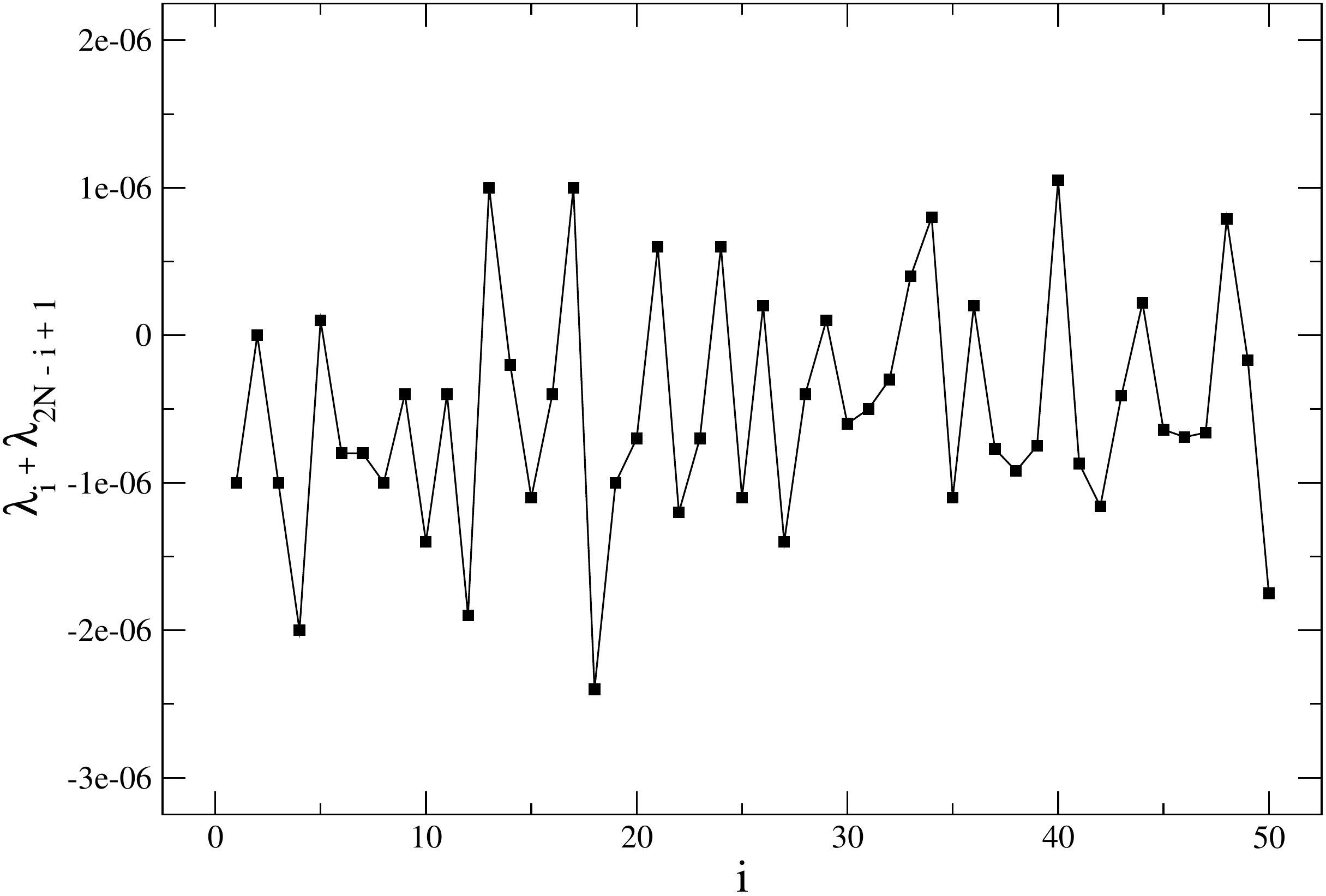}
\end{minipage}
\caption{$N = 50$, $e \approx 0.\,55$. (Left) Lyapunov spectrum ordered according to the expression \eqref{eq.LS}. The abscissa is the index $i = 1,2,\dots,100$ for exponent $\lambda_i$. (Right) Test of relation \eqref{eq.LS_H} to assess how well the Hamiltonian structure of the dynamics is preserved numerically, for each pair of exponents $(i, 2N+1-i)$.} 
		 

\label{fig.LS}
\end{center} 
\end{figure}

Figure \ref{fig.LS} displays a full Lyapunov spectrum for the cosHMF model 
with $N = 50$ particles (that means 100 exponent values). We check also whether our numerical method preserves the symplectic nature of the dynamics by plotting the sum $\lambda_{\imath} + \lambda_{2N - \imath + 1}$ which must vanish according to \eqref{eq.LS_H}.

\begin{figure}[h!]
\centering
\includegraphics[scale=0.49]{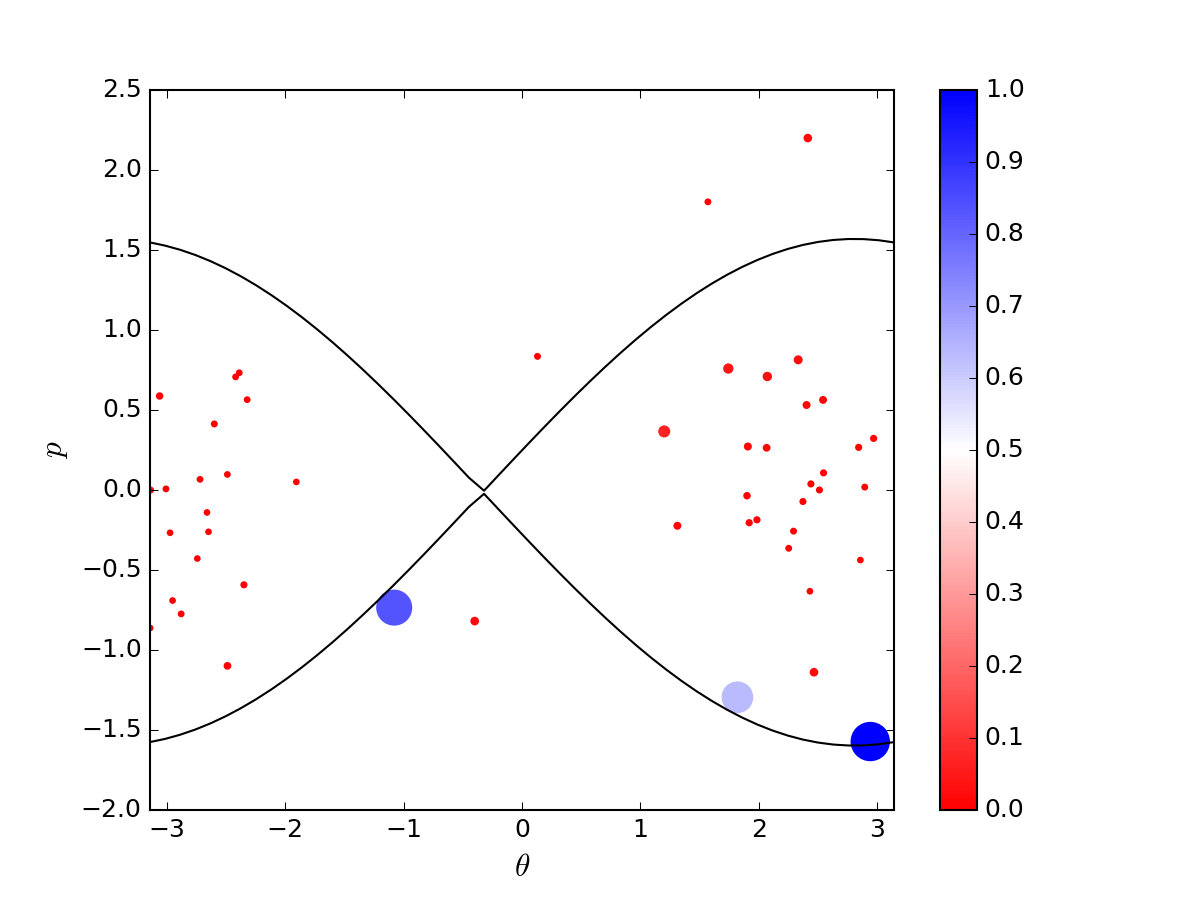}
\caption{$N = 50$, $e \approx 0.\,55$. Snapshot of the cosHMF kinetic space write a colour and size scale based on the weight $\delta I_\ell$ for each particle\textsc{\char13}s contribution to the first Lyapunov vector. Colour scale is relative while the size scale is absolute, proportional to $\delta I_\ell$. The black line is the instantaneous separatrix. In contrast with figures \ref{deltaI_e}, \ref{deltaI_lognormal} and \ref{fig:deltaI_ls}, 
here the abscissa is the particle position $\theta$, rather than its relative position $\theta - \varphi$  with respect to the magnetisation angle.}
\label{deltaI_phase_space}
\end{figure}

Figure \ref{deltaI_phase_space} displays a snapshot of the kinetic space for $N = 50$ particles in the subcritical region ($e \approx 0.\,55$), 
using a colour scale and size scale depending on the value of our indicator $\delta I_\ell$. 
The continuous black line is the instantaneous separatrix, 
with two branches in the typical form of a cat's eye given by (\ref{eq.separ}). 
These two branches meet at the instantaneous ``X point'' associated with the unstable equilibrium 
of the one-particle Hamiltonian (\ref{eq.h1}). 

The colours provide a relative scale of $\delta I_\ell$, 
in which the lowest value is indicated by red and the highest by blue. 
The mark size is used as absolute scale, proportional to $\delta I$. 
This figure shows that most of the $\delta I_\ell$ values are practically null (less than $10^{-5}$),
with few dominant weights located near the separatrix. 
In this equilibrium state, while particles move due to the dynamics, they essentially remain in the same region, 
so that the same particles come recurrently close to the cat's eye, where chaotic motion is prevalent.  

\begin{figure}[!h]
  \begin{center}
  \begin{minipage}[b]{0.49\linewidth}
    \includegraphics[width=\linewidth]{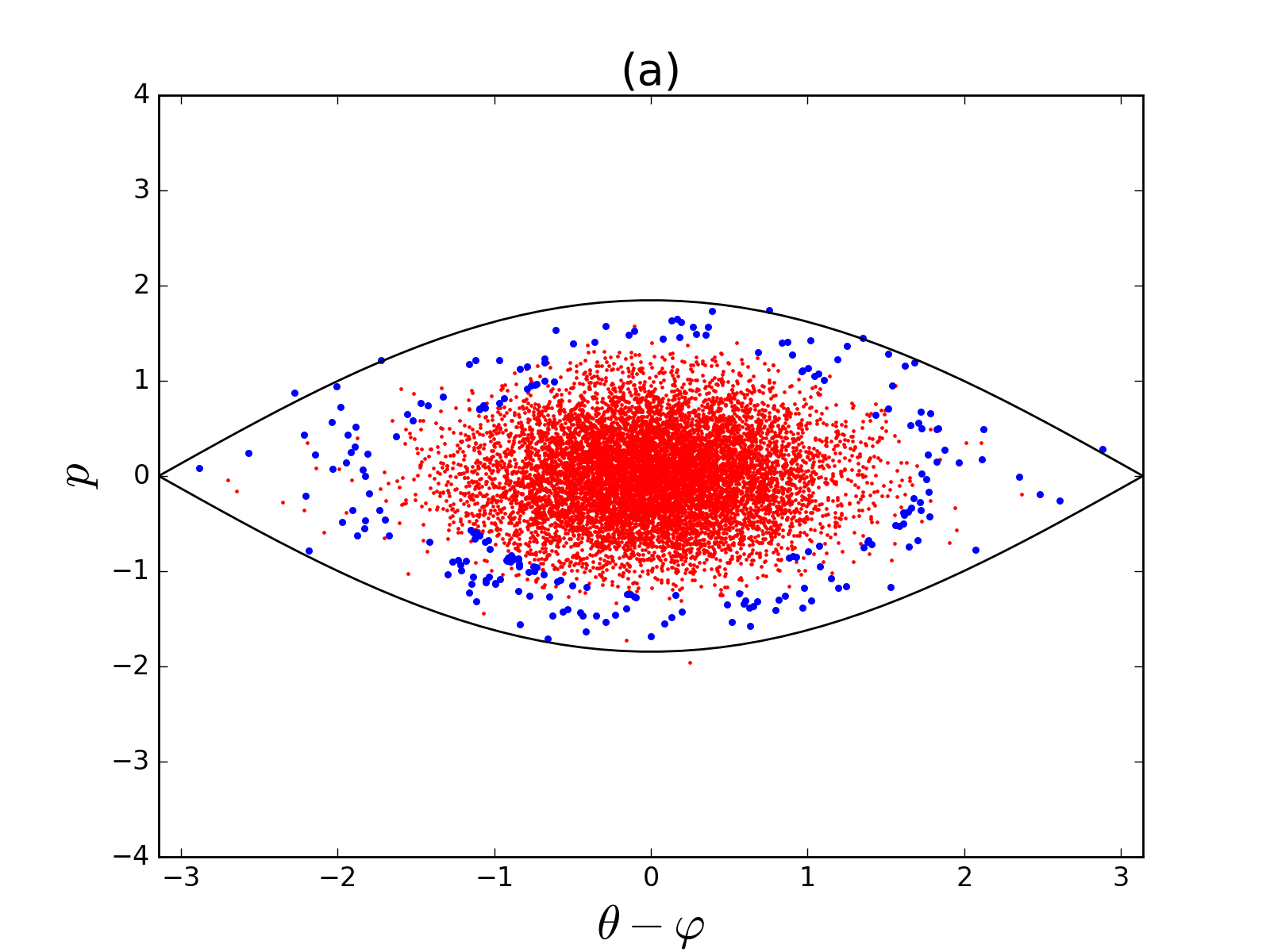}
  \end{minipage} \hfill
  \begin{minipage}[b]{0.49\linewidth}
    \includegraphics[width=\linewidth]{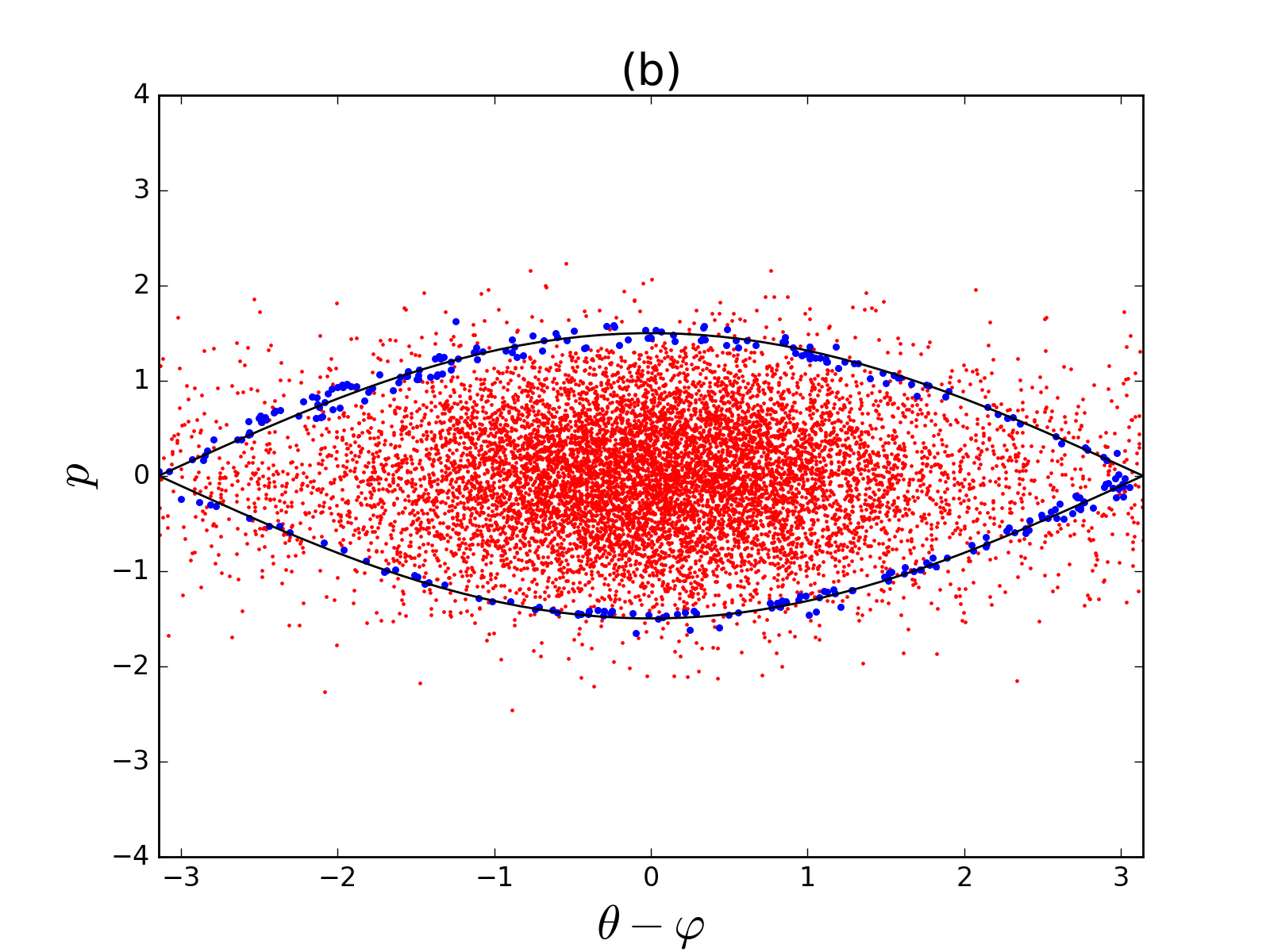}
  \end{minipage} \hfill
  \begin{minipage}[b]{0.49\linewidth}
    \includegraphics[width=\linewidth]{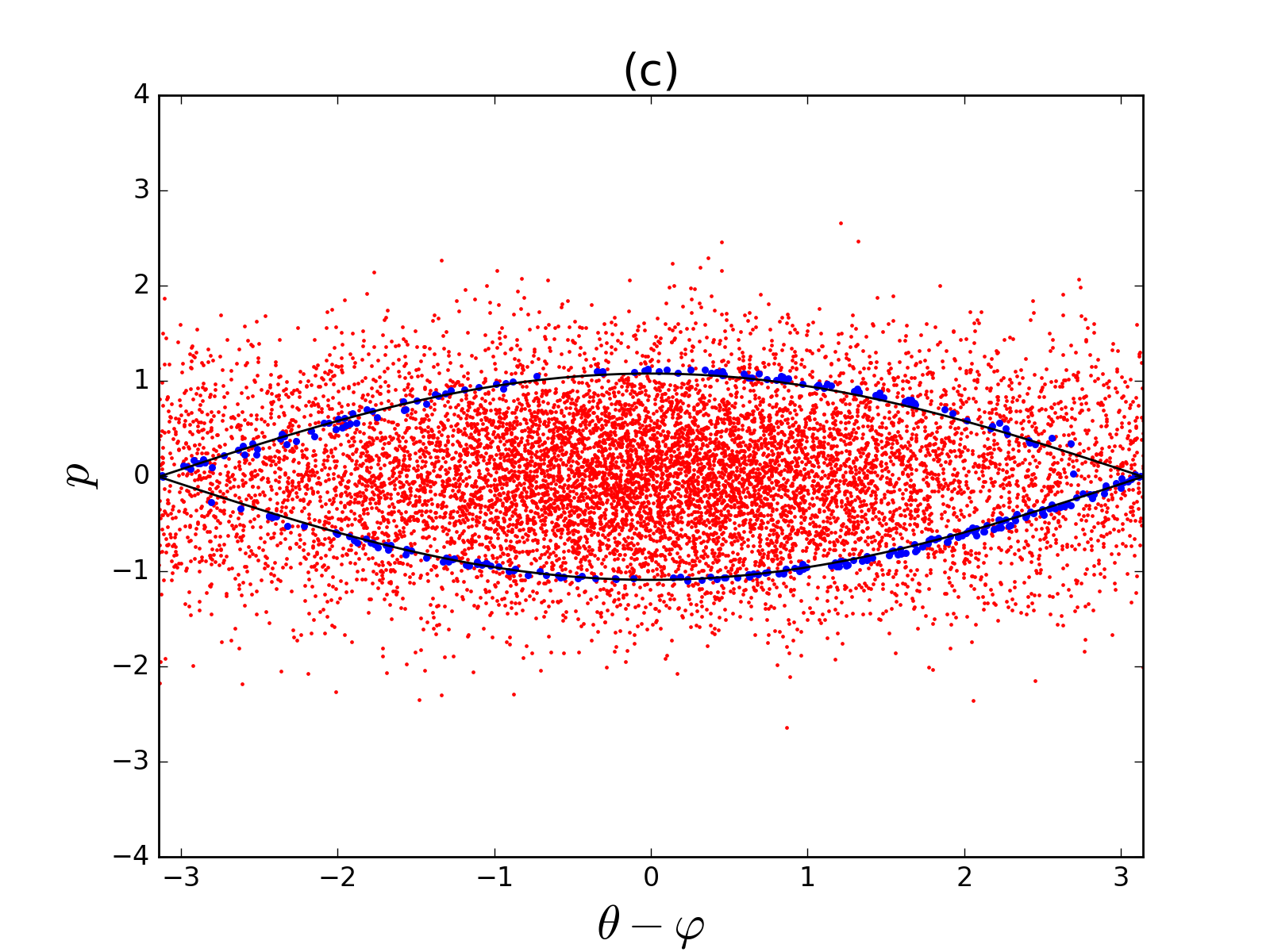}
  \end{minipage} \hfill
  \begin{minipage}[b]{0.49\linewidth}
    \includegraphics[width=\linewidth]{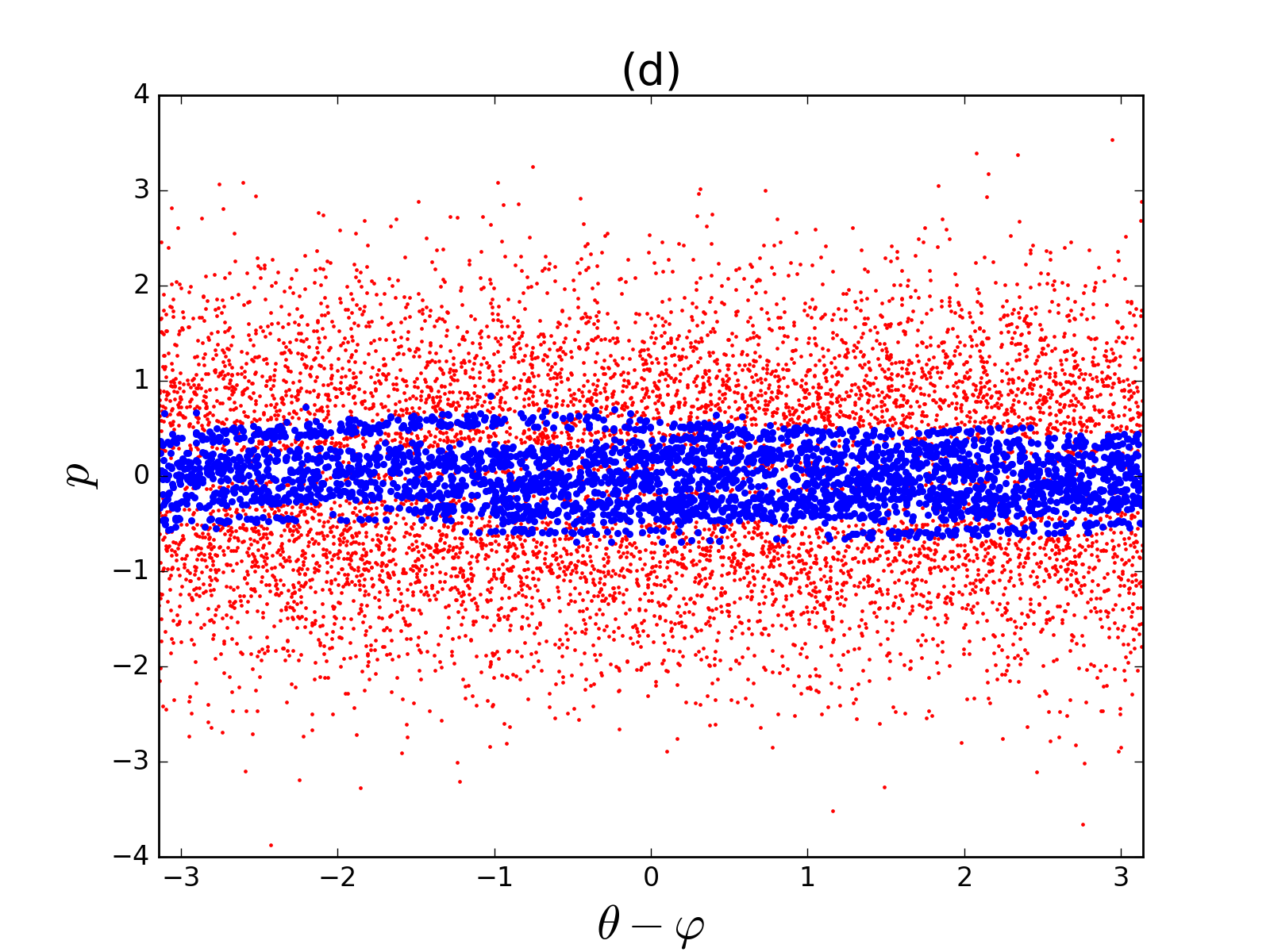}
  \end{minipage}
  \caption{$N = 10^4$. Snapshot of the cosHMF kinetic space for different energy values 
  ($e_{\rm a} \approx 0.\,25$, $e_{\rm b} \approx 0.\,55$, $e_{\rm c} \approx 0.\,70$ and $e_{\rm d} \approx 1.\,00$). 
  Blue dots represent particles for which $\delta I_\ell > 10^{-5}$. 
  These values are located next to the separatrix for subcritical energy cases (panels (a), (b) and (c)), 
  and around $p = 0$ in panel (d), corresponding a disordered phase.}
\label{deltaI_e}
\end{center}
\end{figure}

A similar pattern is also observed for other values of energy.
Figure \ref{deltaI_e} displays representations of the cosHMF kinetic space 
for energies $e_{\rm a} \approx 0.\,25$, $e_{\rm b} \approx 0.\,55$, $e_{\rm c} \approx 0.\,70$ and $e_{\rm d} \approx 1.\,00$. 
The particles highlighted in blue are the particles for which $\delta I_\ell \geq 10^{-5}$.
We clearly see the correlation of the higher values of $\delta I$ with 
delimitant regions of the kinetic space. 
In the subcritical cases (b)(c), the largest values of $\delta I$ are located around the separatrix, 
between bounded and free particles. 
Case (a) corresponds to a confined regime where the particles do not have enough energy to escape from their self-consistent potential well. Nevertheless, the largest values of $\delta I$ are situated near the separatrix. 
With the energy $e_{\rm d} > e^\ast$, the system evolved to a disordered phase 
in which the magnetisation vanishes in the thermodynamic limit and therefore the separatrix is not well defined~; in this case, our numerical results indicate the largest $\delta I_\ell$ values distributed near $p=0$, separating positive and negative velocity states.      

Actually, even when the thermodynamic limit predicts the order parameter $M = 0$ at equilibrium,
the instantaneous value of the numerical output $\vec M$ for an individual realisation is 
of the order of $N^{-1/2}$ by the central limit theorem, 
what defines an instantaneous cat's eye with a width of the order of $N^{-1/4}$.
Particles with a velocity of the order of unity will not be sensitive to this cat's eye 
(though they would slowly diffuse in velocity \cite{ribeiro}, irrespectively of the cat's eye), 
but a particle with an O($N^{-1/4}$) velocity will come close enough to the instantaneous X point, 
and such a particle will have the opportunity to pass the X point (if its energy is high enough) 
or to reverse its velocity (if its energy is too low to pass). 
This will generate a dichotomy on the slow particle behaviour, making it most sensitive to chaos. 

Moreover, when this particle returns close to the X point (after a time of order O($N^{1/4} \ln N$)
as the period of the pendulum diverges near the separatrix), 
the X point and the separatrix will have moved, because $\vec M$ itself fluctuates. 
Hence the particle may have crossed the fluctuating separatrix between its two passages near the X point,
making the new (passing / reversal) dichotomy somewhat decorrelated from the previous one. 
The velocity of slow particles then approaches a random pattern,
similar to the one observed in the slow relaxation of a quasi-stationary state to thermal equilibrium 
for the wave-particle model closely analogous to the cosHMF \cite{firpo-poleni}, 
and in the Chirikov-Taylor standard map \cite{misguich98}.

\begin{figure}[!h]
  \begin{center}
  \begin{minipage}[b]{0.49\linewidth}
    \includegraphics[width=\linewidth]{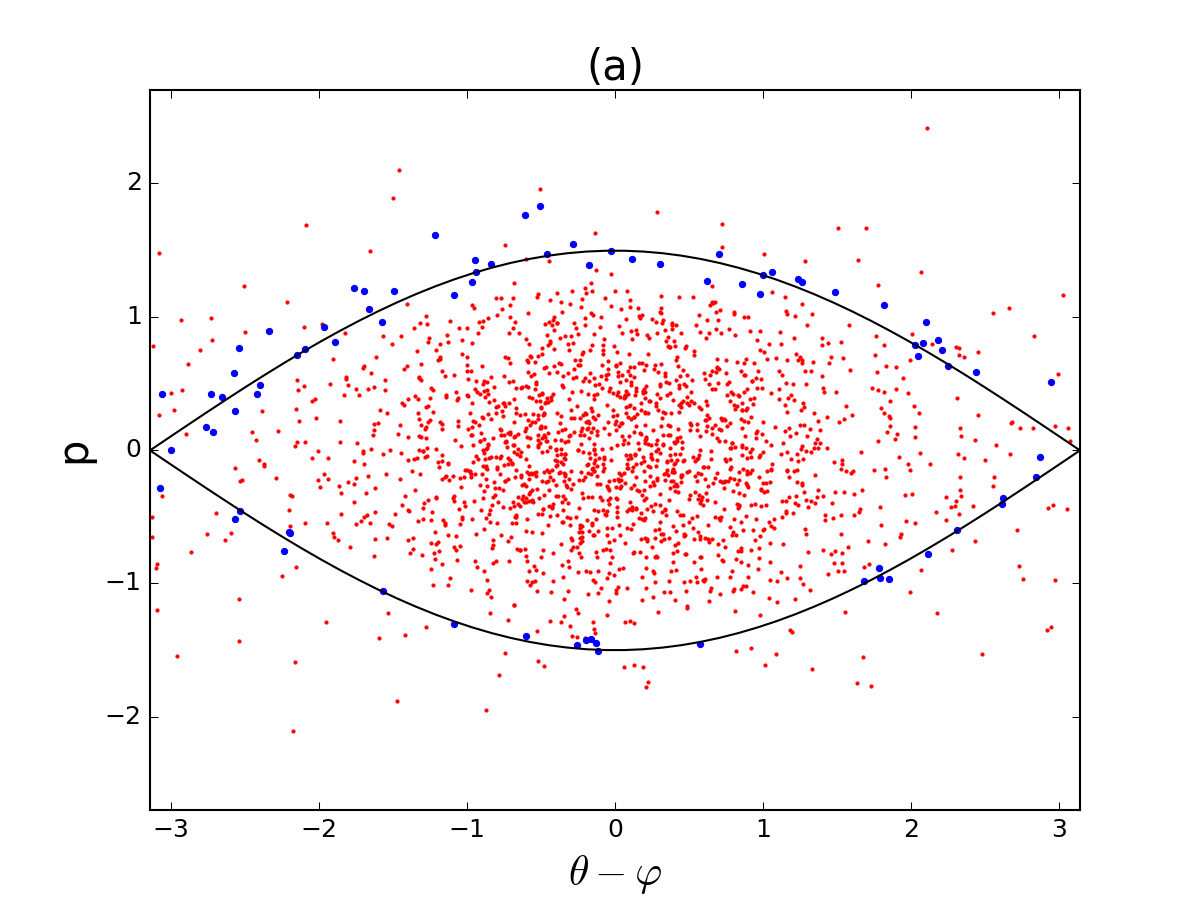}
  \end{minipage} \hfill
  \begin{minipage}[b]{0.49\linewidth}
    \includegraphics[width=\linewidth]{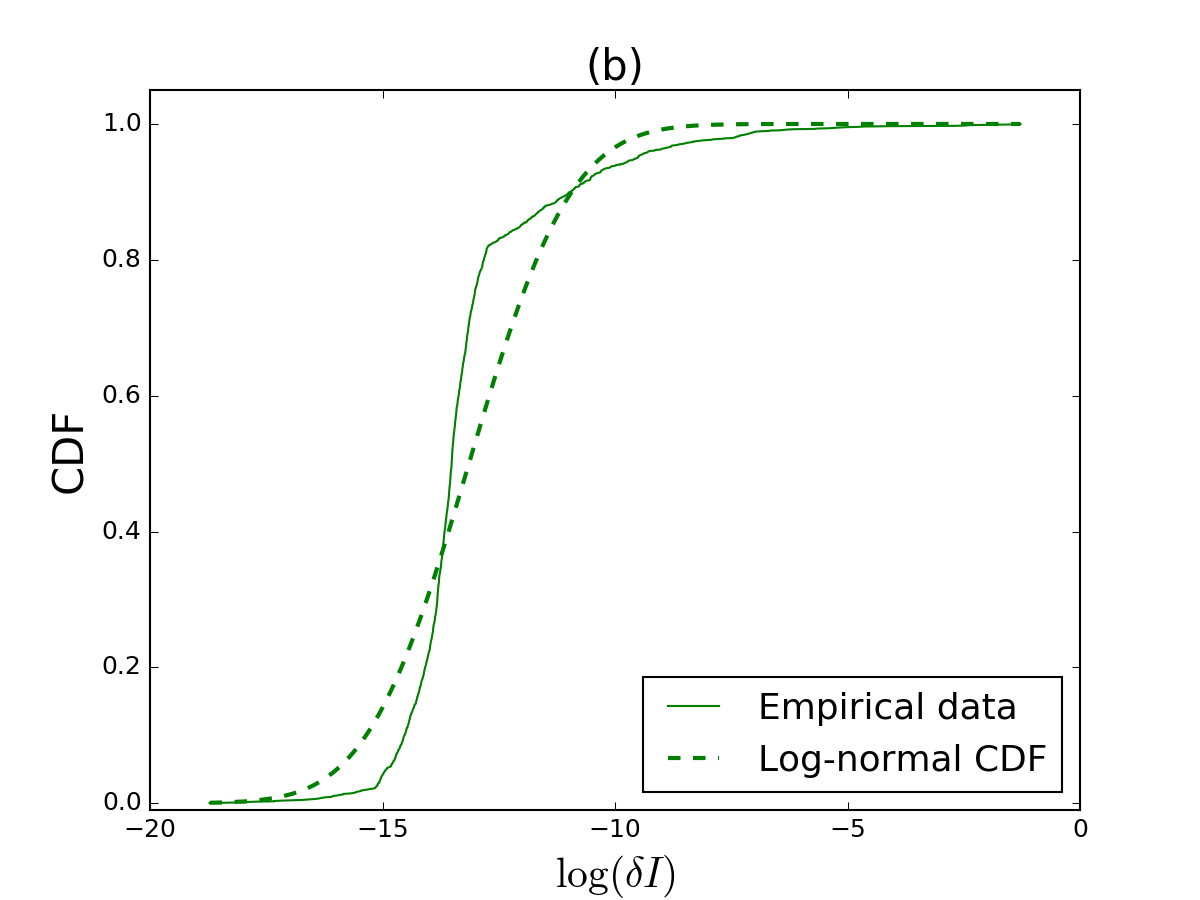}
  \end{minipage} \hfill
  \begin{minipage}[b]{0.49\linewidth}
    \includegraphics[width=\linewidth]{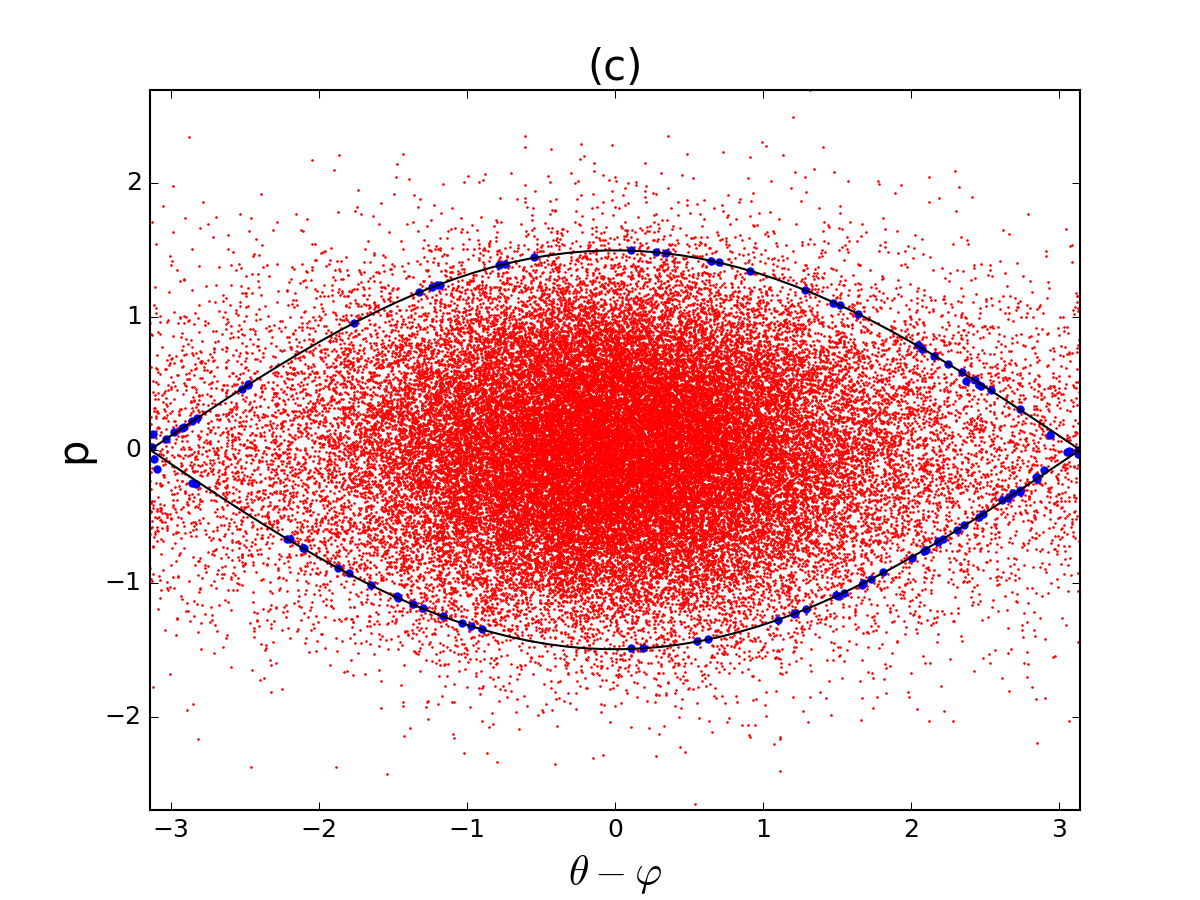}
  \end{minipage} \hfill
  \begin{minipage}[b]{0.49\linewidth}
    \includegraphics[width=\linewidth]{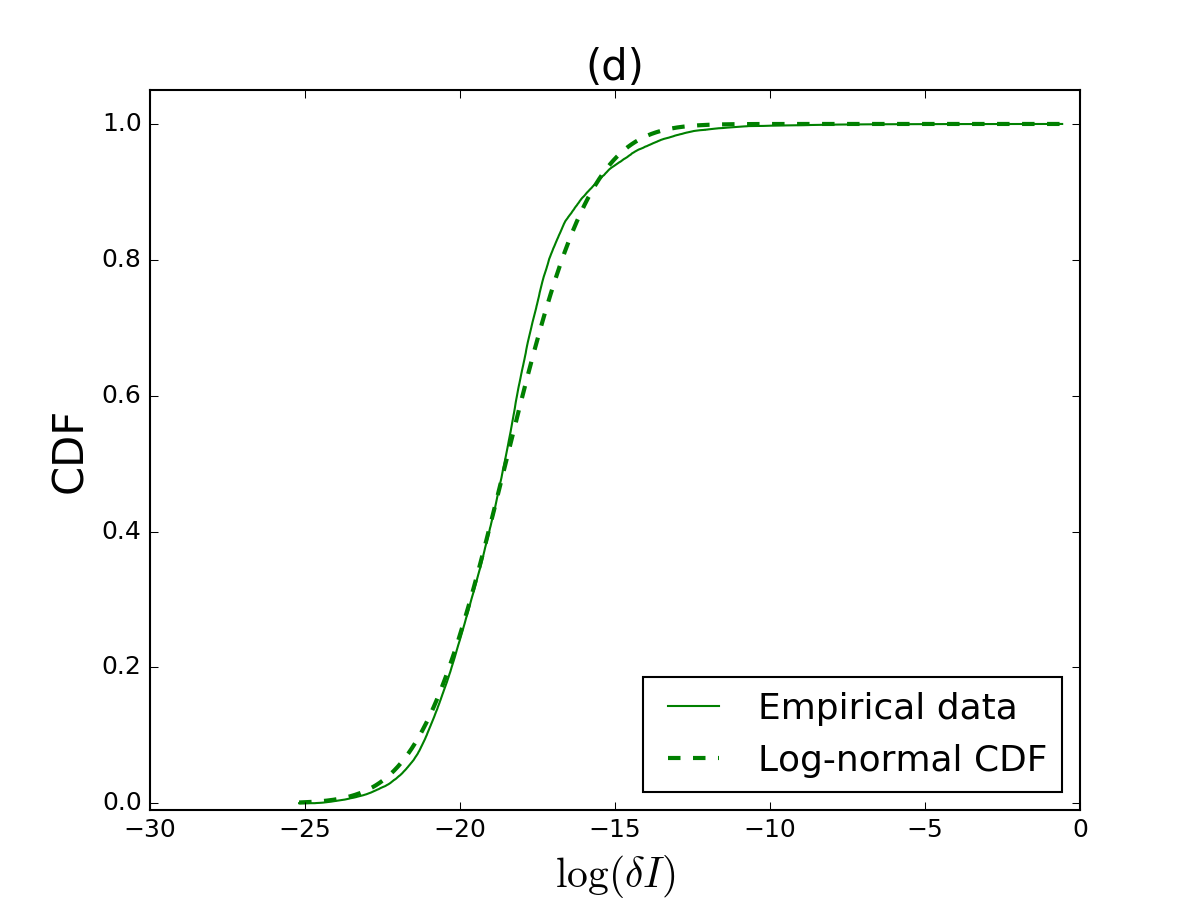}
  \end{minipage}
  \caption{$N_{\rm a} = N_{\rm b} = 2 \times 10^3$, $N_{\rm c} = N_{\rm d} = 5 \times 10^4$, $e \approx 0.\,55$. (Left) Kinetic space with blue dots representing particles for which $\delta I_\ell > 10^{-4}$. (Right) Comparison between cumulative distribution function of  $\log{(\delta I_\ell)}$ and log-normal distribution. Panels (a) and (b) correspond the same case, involving $2 \times 10^3$ particles, as well as (c) and (d) that refer to $5 \times 10^4$ particles.}
\label{deltaI_lognormal}
\end{center}
\end{figure}

Finally, we discuss the distribution of the weights $\delta I_\ell$. 
A rough understanding of this distribution in a quasi-stationary regime with a mildly varying separatrix 
may start from the idea that the one-particle dynamics \eqref{eq.dthetadot}-\eqref{eq.dpdot} in tangent space $(\delta \theta_\ell, \delta p_\ell)$ 
results from multiplying this vector with Jacobian matrices 
generated by the corresponding particle trajectory $(\theta_\ell(t), p_\ell (t))$ and by the successive magnetisations $\vec M(t)$. 
To the extent that the dynamics exhibits chaos, one may expect the successive stretchings 
to rescale the vector $(\delta \theta_\ell, \delta p_\ell)$ more or less independently,
so that $\ln \| (\delta \theta_\ell (t), \delta p_\ell (t)) \| / \| (\delta \theta_\ell (0), \delta p_\ell (0)) \|$ may obey a central limit theorem. 
This would suggest that the observed weights $\delta I_\ell$ could obey a log-normal distribution. 
Although the argument is oversimplistic, it is compatible with the cumulative distributions observed on Figure~\ref{deltaI_lognormal}.

\begin{figure}[!h]
  \begin{center}
  \begin{minipage}[b]{0.49\linewidth}
    \includegraphics[width=\linewidth]{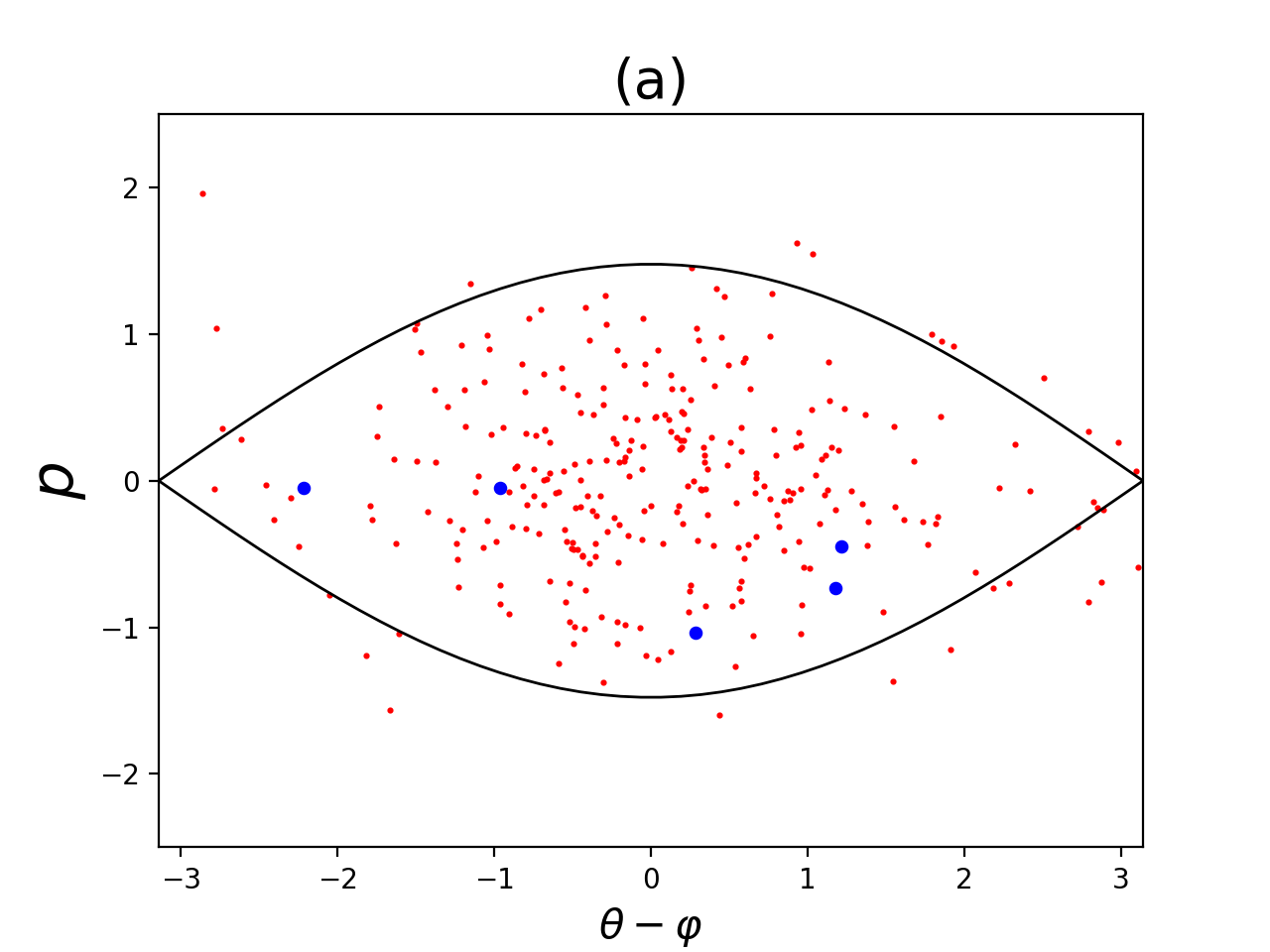}
  \end{minipage} \hfill
  \begin{minipage}[b]{0.49\linewidth}
    \includegraphics[width=\linewidth]{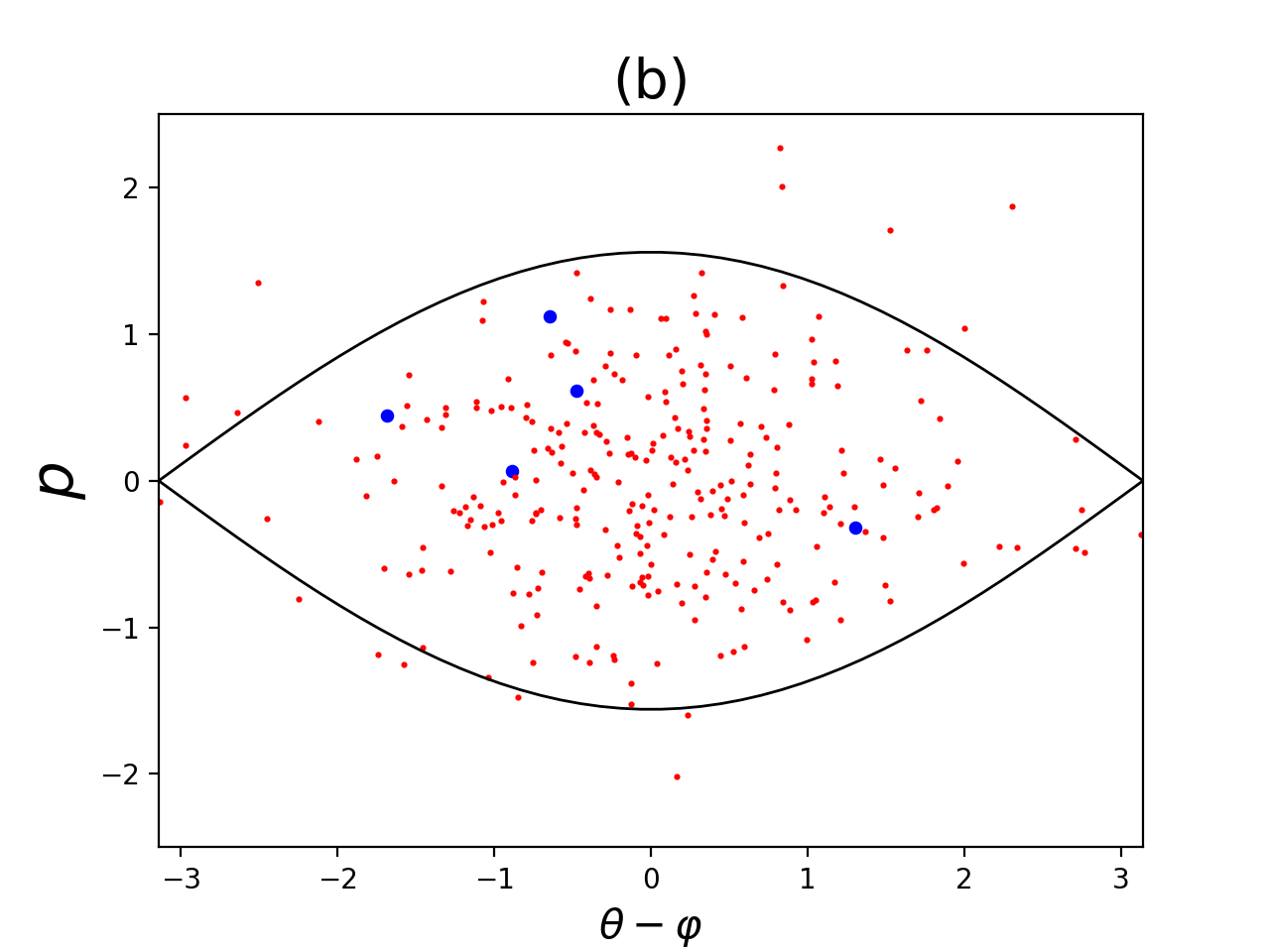}
  \end{minipage} \hfill
  \begin{minipage}[b]{0.49\linewidth}
    \includegraphics[width=\linewidth]{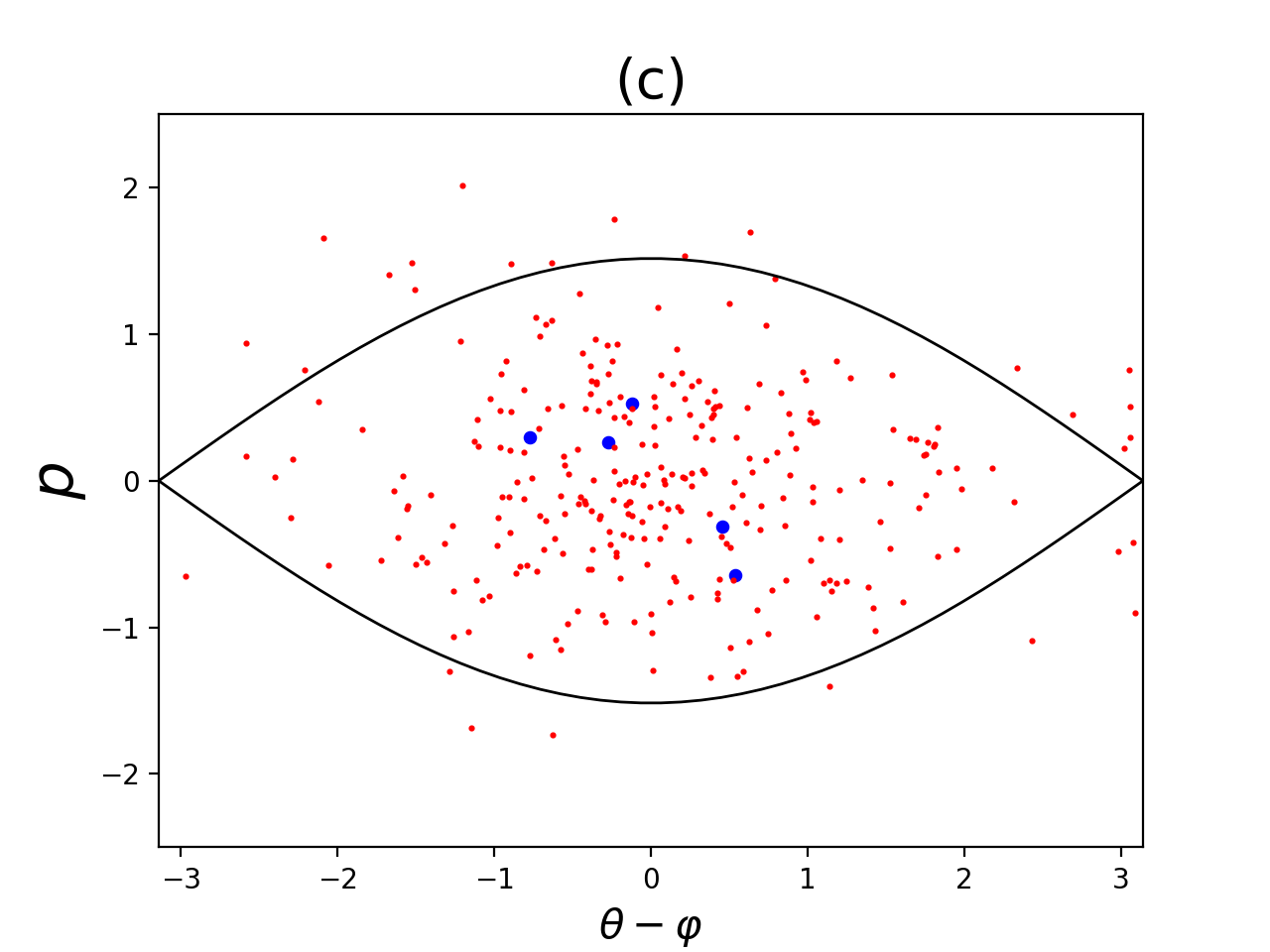}
  \end{minipage} \hfill
  \begin{minipage}[b]{0.49\linewidth}
    \includegraphics[width=\linewidth]{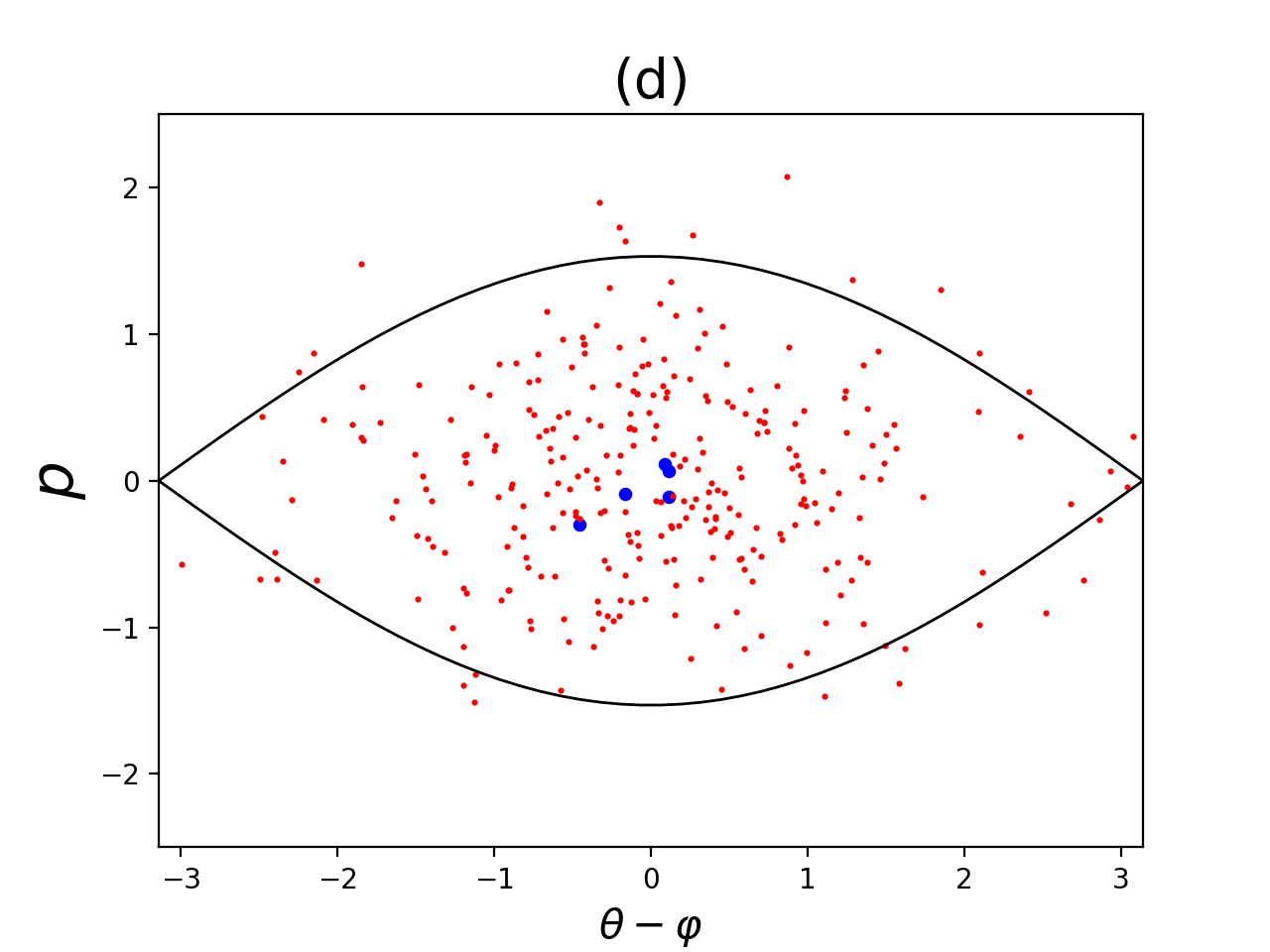}
  \end{minipage}
  \caption{$N_{\rm a} = N_{\rm b} = N_{\rm c} = N_{\rm d} = 3 \times 10^2$, $e 	\approx 0.\,55$. 
          Kinetic space where the blue dots highlight the particles with the five largest values of $\delta I$, 
          for the Lyapunov vectors associated with 
          (a) $\lambda_{75}$, (b) $\lambda_{150}$, (c) $\lambda_{225}$ and (d) $\lambda_{300}$.}
\label{fig:deltaI_ls}
\end{center}
\end{figure}

We also use the weight $\delta I$ to identify the particles carrying most of the contribution 
to the Lyapunov vectors associated with other exponents. 
In Figure \ref{fig:deltaI_ls}, 
for the same energy as in figures \ref{deltaI_phase_space}, \ref{deltaI_e}b and \ref{deltaI_lognormal}, 
we highlight the vectors associated with index $kN/4$ ($1 \leq k \leq 4$), 
so that panel \ref{fig:deltaI_ls}d corresponds to a null Lyapunov exponent. 
In agreement with intuition, the smaller exponents give more weight to particles deeper in the potential well. 

\begin{figure}[!h]
  \begin{center}
  \begin{minipage}[b]{0.49\linewidth}
    \includegraphics[width=\linewidth]{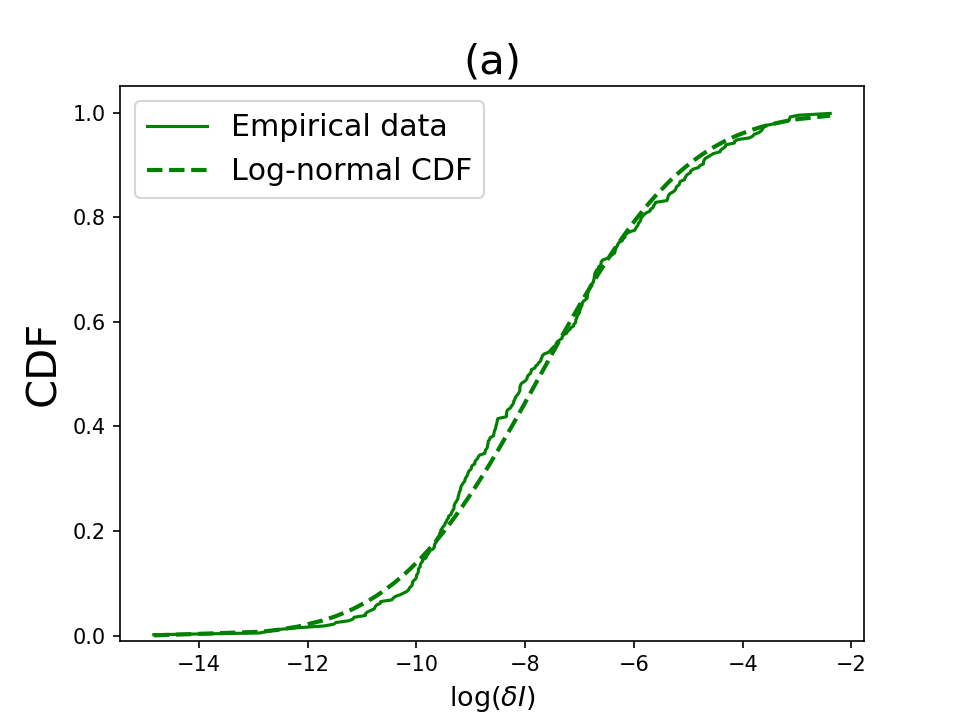}
  \end{minipage} \hfill
  \begin{minipage}[b]{0.49\linewidth}
    \includegraphics[width=\linewidth]{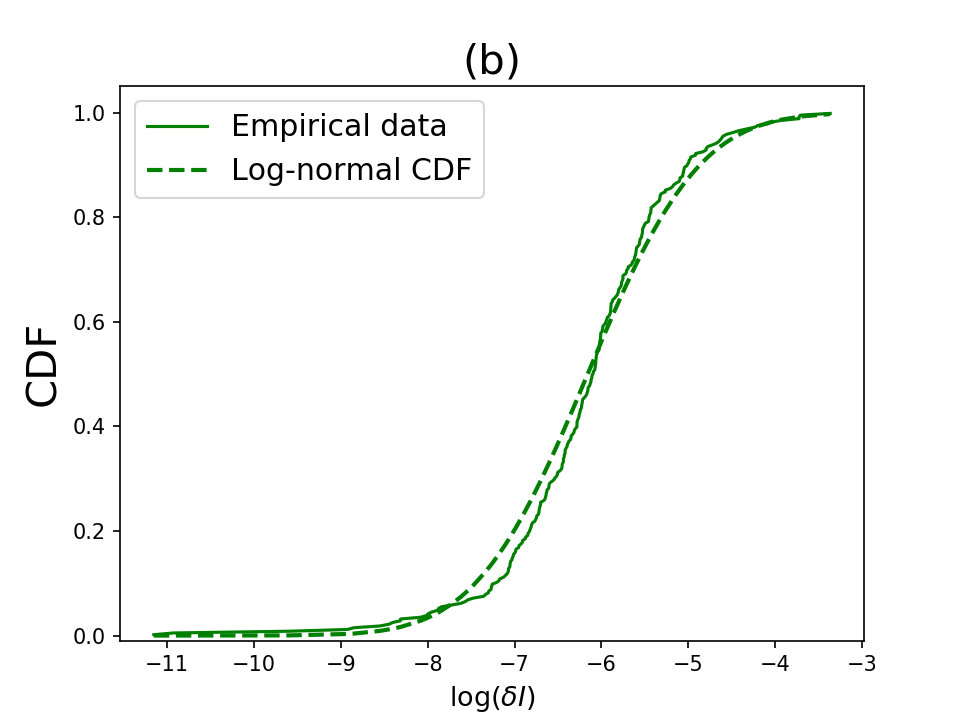}
  \end{minipage} \hfill
  \begin{minipage}[b]{0.49\linewidth}
    \includegraphics[width=\linewidth]{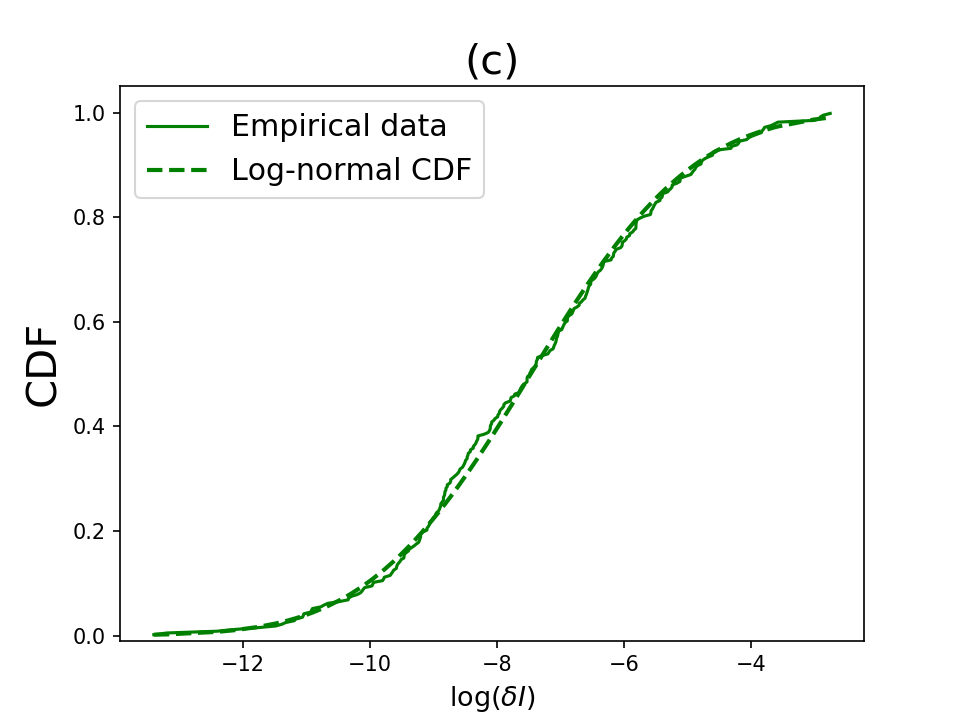}
  \end{minipage} \hfill
  \begin{minipage}[b]{0.49\linewidth}
    \includegraphics[width=\linewidth]{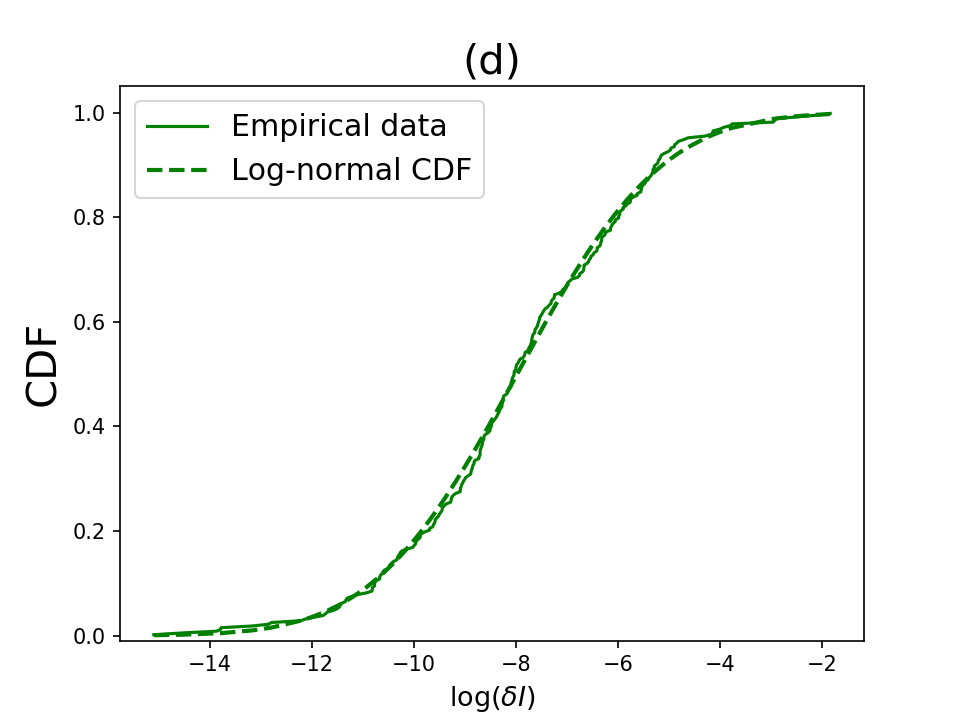}
  \end{minipage}
  \caption{$N_{\rm a} = N_{\rm b} = N_{\rm c} = N_{\rm d} = 3 \times 10^2$, $e \approx 0.\,55$. Comparison between cumulative distribution function of $\delta I_\ell$ and log-normal distribution. Panels correspond to those of figure~\ref{fig:deltaI_ls}.
  }
\label{deltaI_ls_cdf}
\end{center}
\end{figure}

The distribution of weights was also considered for the first Lyapunov vector by Ginelli \emph{et al.}~\cite{ginelli11},
who found a scaling $\sqrt{\delta I_\ell} \sim \ell^{-1}$. As their work focused on the largest Lyapunov exponent and its vector, 
they do not consider higher exponents and associated vectors. 
The localisation of Lyapunov vectors is also discussed by Bosetti and Posch~\cite{bosetti14} 
and Taniguchi and Morris~\cite{taniguchi2003B}, 
using a global measure of localisation like $N^{-1} \exp(- \sum_\ell \delta I_\ell \ln \delta I_\ell)$. 
Here we consider the detailed distribution of the weights for several Lyapunov vectors. 
If these weights had a power-law distribution, say $\delta I_\ell \approx C \ell^{-k}$ 
(where $k = 2$ would correspond to Figure~\ref{deltaI_e} of Ref.~\cite{ginelli11}), 
then the cumulative distribution function of the weights would scale like $(\delta I)^{1 - 1/k}$.

\begin{figure}[!h]
  \begin{center}
  \begin{minipage}[b]{0.49\linewidth}
    \includegraphics[width=\linewidth]{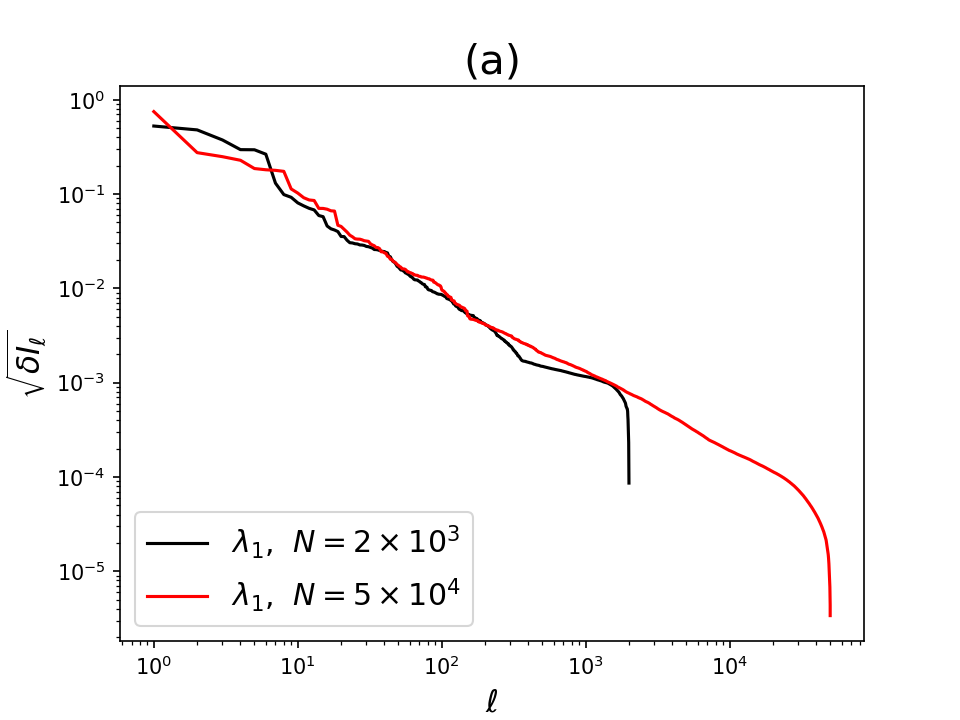}
  \end{minipage} \hfill
  \begin{minipage}[b]{0.49\linewidth}
    \includegraphics[width=\linewidth]{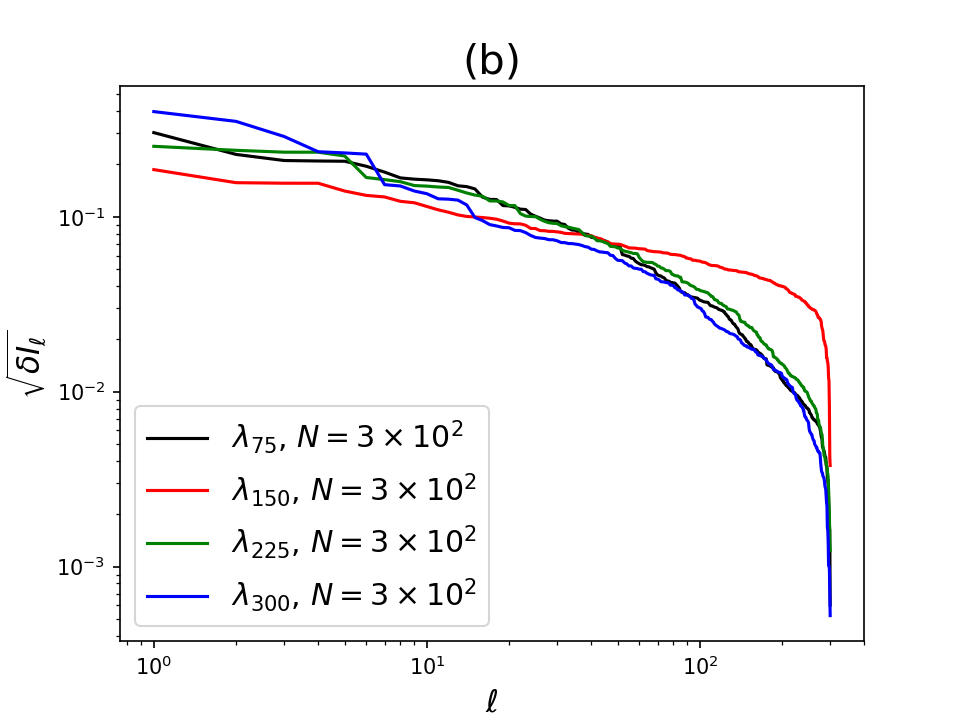}
  \end{minipage} \hfill
\caption{$\sqrt{\delta I_\ell}$ vs $\ell$, $e \approx 0.\,55$. $\sqrt{\delta I_\ell}$ sorted in decreasing order. Panel (a) corresponds to Figures \ref{deltaI_lognormal} and panel (b) to Figures~\ref{fig:deltaI_ls} and \ref{deltaI_ls_cdf} .} 
\label{final_1}
\end{center}
\end{figure}

\begin{figure}[!h]
  \begin{center}
  \begin{minipage}[b]{0.49\linewidth}
    \includegraphics[width=\linewidth]{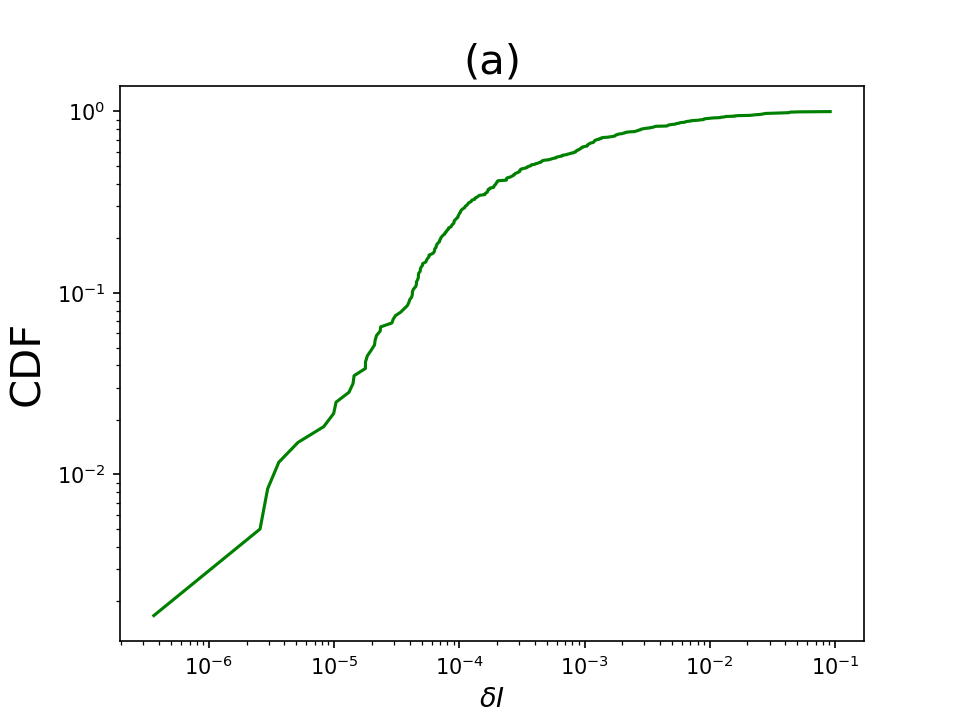}
  \end{minipage} \hfill
  \begin{minipage}[b]{0.49\linewidth}
    \includegraphics[width=\linewidth]{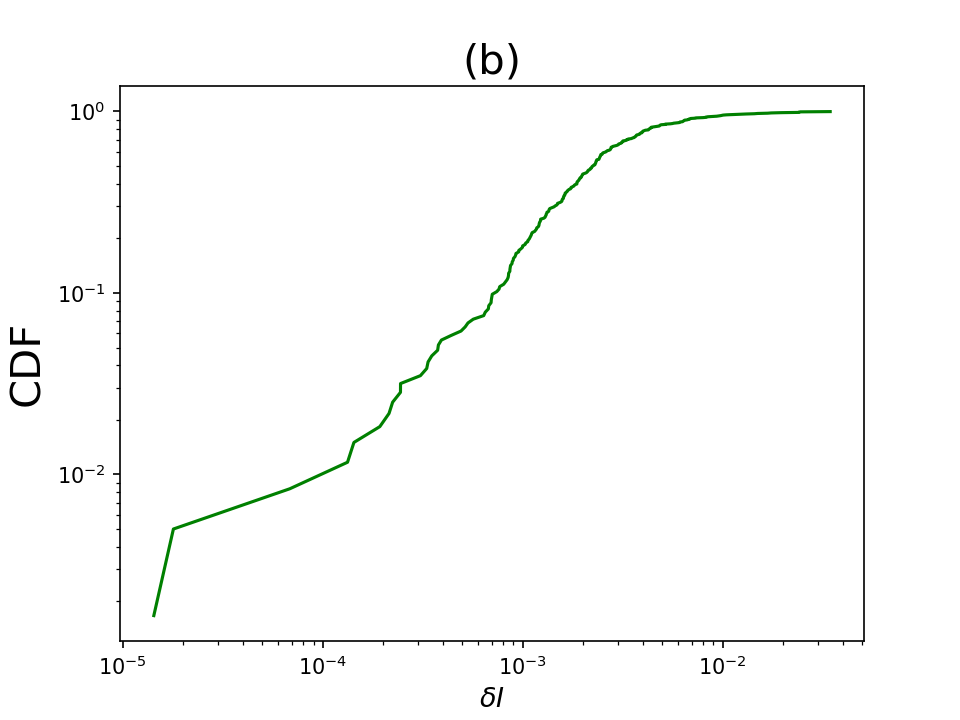}
  \end{minipage} \hfill
  \begin{minipage}[b]{0.49\linewidth}
    \includegraphics[width=\linewidth]{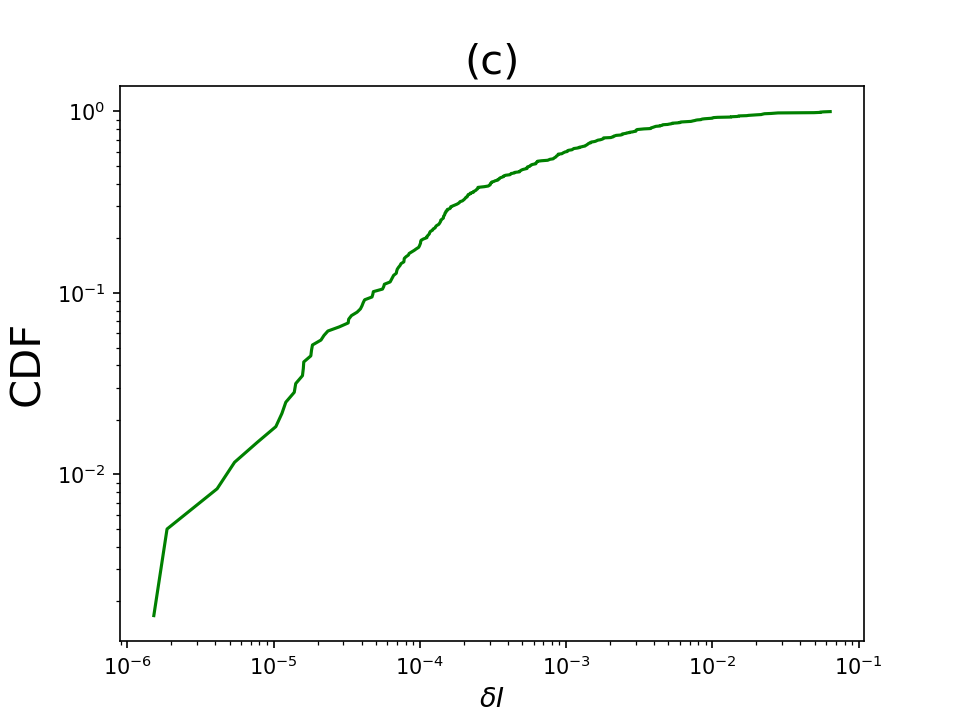}
  \end{minipage} \hfill
  \begin{minipage}[b]{0.49\linewidth}
    \includegraphics[width=\linewidth]{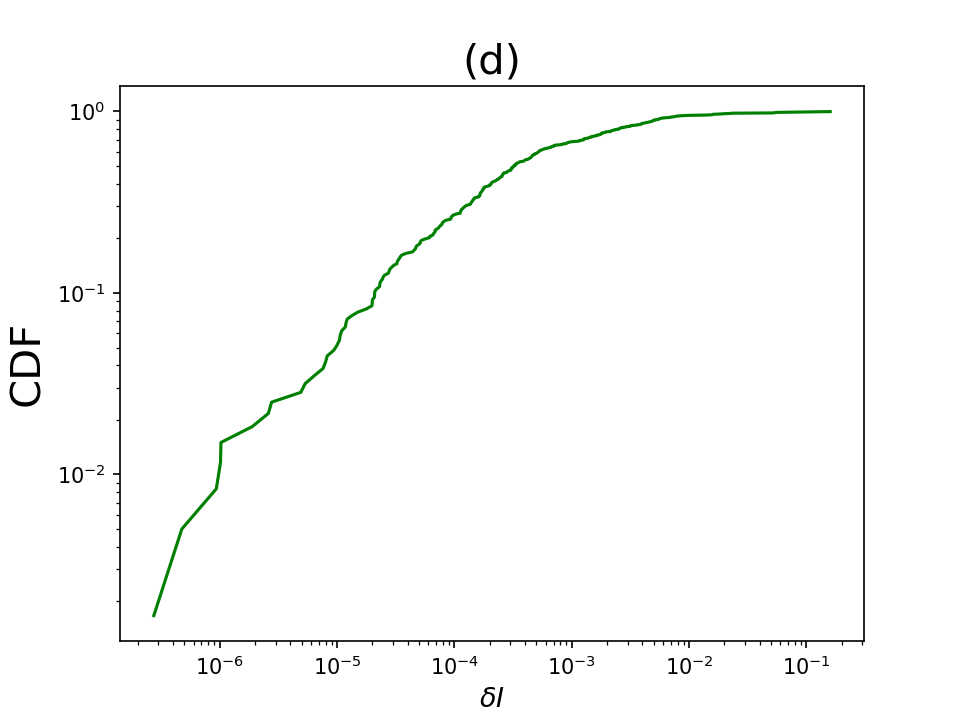}
  \end{minipage}
\caption{Cumulative distribution function of $\delta I_\ell$ : a power-law distribution $\delta I_\ell \sim \ell^{-k}$ would appear as a straight line with slope $1 - 1/k$. 
              Panels correspond to figures~\ref{fig:deltaI_ls} and \ref{deltaI_ls_cdf}.  }
\label{deltaI_ls_cdf2}
\end{center}
\end{figure}

Figure~\ref{final_1} presents the values of $\sqrt{\delta I_\ell}$ in function of its index $\ell$. As noted in Ref.~\cite{ginelli11} for the Lyapunov vectors associated with $\lambda_1$, we have $\sqrt{\delta I_\ell} \sim \ell^{-1}$. The observed distributions for other Lyapunov vectors of the spectrum, Figure~\ref{final_1}(b), do not completely rule out the power-law scaling, but do not fully support it either, as well as the cumulative distributions displayed in Figure~\ref{deltaI_ls_cdf2}.

\section{Concluding remarks}
\label{sec.conc}

We associate with each particle in an $N$-body system a weight $\delta I_\ell$ measuring their respective contribution to the chaoticity of the dynamics, 
in terms of the components of the first Lyapunov vector, in a macroscopically stationary regime. 
While our numerical determination of $\delta I_\ell$ derives from the Gram-Schmidt procedure for evolving the tangent dynamics \eqref{eq.linear}, 
this weight refers only to the largest exponent, 
and therefore corresponds to the covariant Lyapunov vector as well as to the ``orthogonal'' Lyapunov vector \cite{posch}. 
It is thus genuinely intrinsic to the dynamics.  

We applied this procedure to molecular dynamics simulations of the paradigmatic cosHMF model. 
We checked our calculations accuracy by computing the whole Lyapunov spectrum. 
According to the weight $\delta I$, the most chaotic regions in the kinetic space 
turn out to be, for the cosHMF model at subcritical values of energy ($e < e^{\star}$), the vicinity of the separatrix, 
in agreement with previous results~\cite{ginelli11}. 
High energy ($e > e^{\star}$) implies a gaslike regime for the particle motion, 
in which the largest $\delta I$ values are distributed around the particles for which $p=0$ : 
this domain is also the one where instantaneous one-particle dynamics 
shows a small, fluctuating separatrix \cite{ettoumi-firpo,ribeiro}.  

We also discuss the distribution of weights among particles for Lyapunov vectors associated with higher-index exponents. 

Thanks to its simplicity and its robustness, our approach may help in identifying chaotic features of more complex interactions \cite{antoniES}
and in providing a way to interpret the behaviour of these systems with a reduced number of degrees of freedom.

\section*{Acknowledgements}

LHMF thanks the PIIM, CNRS-AMU for their hospitality, Bruno V.\ Ribeiro and Daniel Santos for productive discussions, the remote access computing laboratory of the DF-UFRPE and CIF-UnB for efficient numerical help, 
and CAPES for financial support. TMRF was partially financed by CNPq (Brazil). 
YE enjoyed the hospitality and support from CIFMC/UnB while completing this work. 
It is a pleasure to gratefully acknowledge fruitful discussions with Dominique F.\ Escande 
and members of the teams ``turbulence plasma'' and ``dynamique des syst\`emes complexes'' in Marseille.


\end{document}